\documentclass[showpacs,preprint]{revtex4}

\usepackage{amsmath,amssymb,mathrsfs}
\usepackage{latexsym}
\usepackage{graphicx,psfrag} 

\newcommand{\bs}[1]{\boldsymbol{#1}}

\newcommand{\comm}[2]{\left[#1,#2\right]}

\newcommand{\half}{$\frac{1}{2}$ }
\newcommand{\vac}{\left|\,0\,\right\rangle}
\newcommand{\ket}[1]{\left|#1\right\rangle}

\newtheorem{theo}{Theorem}

\def\bd{\begin{displaymath}}
\def\ed{\end{displaymath}}
\def\be{\begin{equation}}
\def\ee{\end{equation}}
\def\bea{\begin{eqnarray}}
\def\eea{\end{eqnarray}}
\def\bi{\begin{itemize}}
\def\ei{\end{itemize}}
\def\bn{\begin{enumerate}}
\def\en{\end{enumerate}}

\def\ie{\emph{i.e.,\ }}
\def\eg{\emph{e.g.\ }}

\def\ho{{\text{ho}}}
\def\b{{\text{b}}}
\def\r{{\text{r}}}
\def\g{{\text{g}}}

\def\bb{\text{bb}}
\def\rr{\text{rr}}

\def\SU(3){\text{SU(3)}}

\allowdisplaybreaks[1]
\begin{document}
\title{Charge excitations in SU($\bs{n}$) spin chains:\\
exact results for the 1/$\bs{r}^{\bs{2}}$ model}
\author{Ronny Thomale, Dirk Schuricht\cite{dirk}, and Martin Greiter}
\affiliation{Institut f\"ur Theorie der Kondensierten Materie,\\
  Universit\"at Karlsruhe, Postfach 6980, 76128 Karlsruhe, Germany}
\pagestyle{plain}
\begin{abstract}
  We study the one- and two-holon excitations of the SU(3)
  Kuramoto--Yokoyama model on the level of explicit wave functions,
  and generalize the calculations to the case of SU($n$). We obtain
  the exact energies and the single holon momenta, which we find
  fractionally spaced according to fractional statistics with
  statistical parameter $g=1/n$.  
\end{abstract} 
\maketitle

\section{Introduction}\label{secintro}

Since its discovery in 1988 by Haldane~\cite{Haldane88} and
Shastry~\cite{Shastry88}, the Haldane--Shastry model (HSM) has amply
contributed to our understanding of fractional quantization in
one-dimensional spin chains.  The model provides a framework to
formulate and analyze spinons, the elementary excitations of
one-dimensional spin chains, at the level of explicit wave
functions~\cite{haldane91prl1529,BGLtwospinon}.  In particular, it was
realized through this model that spinons in SU(2) spin chains obey
half-Fermi statistics~\cite{Haldane91prl2}.
Kawakami~\cite{kawakami92prb1005} subsequently generalized the HSM
from SU(2) spins to SU($n$), a model in which the spinon excitations
obey fractional statistics with statistical parameter
$(1-1/n)$~\cite{kuramoto,kato,Schoutens97,BouwknegtSchoutens99,
  Yamamoto-00prl,Yamamoto-00jpsj,Schuricht-06prb}.

The HSM was also generalized by Kuramoto and
Yokoyama~\cite{KuramotoYokoyama} to allow for mobile holes.  The
Kuramoto--Yokoyama Model (KYM) hence contains spin and charge degrees
of freedom, described by spinon and holon
excitations~\cite{ha-94prl2887}, which carry spin \half but no charge
and charge $+1$ but no spin, respectively.  While explicit wave
functions for one-holon states in the SU(2) KYM were known for many
years~\cite{LauglinChia}, the construction of the exact two-holon
states was achieved only recently~\cite{TSG}.  In particular, the
single-holon momenta in these states were found to be shifted by a
fraction of the units $2\pi/N$ appropriate for a chain with $N$ sites,
periodic boundary conditions (PBCs), and a lattice constant set to
unity.  This result was interpreted as a manifestation of half-Fermi
and hence fractional statistics among the holon
excitations~\cite{Greiter06}, thus confirming a conclusion reached by
Ha and Haldane~\cite{ha-94prl2887} using the asymptotic Bethe ansatz,
by Kuramoto and Kato~\cite{kuramoto,kato} from thermodynamics, and by
Arikawa, Yamamoto, Saiga, and Kuramoto~\cite{Arikawa1,Arikawa2} from
the electron addition spectral function of the model.  Like the HSM,
the KYM can be generalized to spin symmetry
SU($n$)~\cite{kawakami92prb1005,ha-94prl2887}.

In this article, we analyze the one-holon and two-holon excitations of
the SU($n$) KYM on the level of explicit wave functions.  The article
is organized as follows.  In Section~\ref{secsu3}, we investigate the
case of SU(3).  We first present the basic properties of the model
including the ground state and the coloron excitations in the absence
of holes, where the SU(3) KYM reduces to the SU(3) HSM studied
previously in a similar framework~\cite{Schuricht-06prb}.  We then
construct the explicit one-holon and two-holon wave functions and
derive the exact energies and single-holon momenta.  In
Section~\ref{secsun}, we generalize the results to SU($n$).  In
particular, we review the basic properties of the ground state and the
SU($n$) spinon excitations before we derive the one-holon and
two-holon wave functions including their energies and momenta.  In
Section~\ref{secfrac}, we interpret our results in terms of free
holons obeying fractional statistics with statistical parameter
$g=1/n$.

\section{SU(3) Kuramoto--Yokoyama Model}\label{secsu3}

\subsection{Hamiltonian}\label{subsecsu3ham}

The SU(3) Kuramoto--Yokoyama model (KYM)~\cite{kawakami92prb1005} is
most conveniently formulated by embedding the one-dimensional chain
with periodic boundary conditions into the complex plane by mapping it
onto the unit circle with the sites located at the complex positions
$\eta_\alpha=\exp\!\left(i\frac{2\pi}{N}\alpha\right)$, where $N$
denotes the number of sites and $\alpha=1,\ldots,N$.  For the SU(3)
case, the sites can be either singly occupied by a fermion with
SU(3) spin or empty.  The Hamiltonian is given by
\begin{equation}
H_{\text{SU(3)}}=-\frac{\pi^2}{N^2}
\sum_{\substack{\alpha,\beta=1 \\\alpha \neq \beta}}^{N}
\frac{P_{\alpha \beta}}{\vert \eta_{\alpha} - \eta_{\beta} \vert^2},
\label{ky1}
\end{equation}
where $P_{\alpha \beta}$ exchanges the configurations on the sites
$\eta_{\alpha}$ and $\eta_{\beta}$ including a minus sign if both are
fermionic. Rewriting \eqref{ky1} in terms of spin and fermion creation
and annihilation operators yields
\begin{equation}
H_{\text{SU(3)}} =  \frac{2\pi^2}{N^2} 
\sum_{\alpha \neq \beta}^N \frac{1}{ | \eta_\alpha - \eta_\beta |^2 } 
P_{\text{G}} \Biggl[-\frac{1}{2} 
\sum_{\sigma=\b,\r,\g}\Bigl(c_{\alpha \sigma}^\dagger 
c_{\beta \sigma}^{\phantom{\dagger}} 
+ c_{\beta \sigma}^\dagger c_{\alpha \sigma}^{\phantom{\dagger}} \Bigr)
+\bs{J}_\alpha \cdot \bs{J}_\beta
-\frac{n_{\alpha}n_{\beta}}{3}+n_{\alpha}-\frac{1}{2}\Biggr] P_{\text{G}},
\label{kysu3}
\end{equation}
\noindent
where we label the SU(3) spin or color index $\sigma$ by the colors
blue ($\b$), red ($\r$), and green ($\g$). The Gutzwiller projector
enforces at most single occupancy on all sites, and is explicitly
given by
\begin{equation}
\label{eq:gwproj}
P_\mathrm{G}=\prod_{\alpha=1}^N(1-n_{\alpha\b}n_{\alpha\r}
-n_{\alpha\b}n_{\alpha\g}-n_{\alpha\r}n_{\alpha\g}
+2n_{\alpha\b}n_{\alpha\r}n_{\alpha\g}),
\end{equation}
where
$n_\alpha=c_{\alpha\b}^{\dagger}c_{\alpha\b}^{\phantom{\dagger}}+
c_{\alpha\r}^{\dagger}c_{\alpha\r}^{\phantom{\dagger}}+
c_{\alpha\g}^{\dagger}c_{\alpha\g}^{\phantom{\dagger}}$ is the charge
occupation operator at site $\eta_\alpha$.  Furthermore, we have
introduced $\bs{J}_{\alpha}=\frac{1}{2}\sum_{\sigma\tau}
c_{\alpha\sigma}^{\dagger} \bs{\lambda}_{\sigma\tau}
c_{\alpha\tau}^{\phantom{\dagger}}$, the eight-dimensional SU(3) spin
vector, where $\bs{\lambda}$ denotes the vector consisting of the
eight Gell-Mann matrices (see App.~\ref{app:conventions}), and
$\sigma$ and $\tau$ are again SU(3) color indices. For all practical
purposes, it is convenient to express the SU(3) spin operators in
terms of colorflip operators $e_{\alpha}^{\sigma \tau} \equiv
c_{\alpha \sigma}^{\dagger}c_{\alpha \tau}^{\phantom{\dagger}}$. The
Hamiltonian \eqref{kysu3} then becomes
\begin{equation}
H_{\text{SU(3)}} =  \frac{2\pi^2}{N^2} 
\sum_{\alpha \neq \beta}^N \frac{1}{ | \eta_\alpha - \eta_\beta |^2 } 
P_{\text{G}} \Biggl[-\frac{1}{2} 
\sum_{\sigma}\Bigl(c_{\alpha \sigma}^\dagger 
c_{\beta \sigma}^{\phantom{\dagger}} 
+ c_{\beta \sigma}^\dagger c_{\alpha \sigma}^{\phantom{\dagger}} \Bigr)
+\frac{1}{2}\sum_{\sigma,\tau}e_{\alpha}^{\sigma \tau}e_{\beta}^{\tau \sigma}
-\frac{n_{\alpha}n_{\beta}}{2}+n_{\alpha}-\frac{1}{2}\Biggr] P_{\text{G}},
\label{kysu32}
\end{equation}
where the color double sum includes terms with $\sigma=\tau$.

The KYM is supersymmetric, \ie the Hamiltonian \eqref{ky1} commutes with the
operators $J^{ab}=\sum_\alpha a_{\alpha a}^{\dagger}a_{\alpha
  b}^{\phantom{\dagger}}$, where $a_{\alpha a}$ denotes the
annihilation operator of a particle of species $a$ ($a$ runs over
color indices as well as empty site) at site $\eta_\alpha$. The
traceless parts of the operators $J^{ab}$ generate the Lie
superalgebra su(1$|$3), which includes in particular the total spin
operators $\bs{J}=\sum_{\alpha=1}^N\bs{J}_\alpha$.  In addition, the
KYM possesses a super-Yangian symmetry~\cite{ha-94prl2887}, which
causes its amenability to rather explicit solution.

\subsection{Vacuum state}\label{subsecsu3vac}

We first review the state containing no excitations, \ie neither
colorons nor holons. This vacuum state is the ground state at one
third filling, where the SU(3) KYM reduces to the SU(3) HSM.  The
vacuum state for $N=3M$ ($M$ integer) is constructed by Gutzwiller
projection of a filled band (or Slater determinant (SD) state)
containing a total of $N$ SU(3) particles obeying Fermi statistics
\begin{equation}
  \ket{\Psi_0}=P_ {\text{G}}\prod_{|q|\le q_{\text{F}}}
  c_{q\text{b}}^\dagger\,c_{q\text{r}}^\dagger\,c_{q\text{g}}^\dagger
  \ket{0}
  \equiv P_{\text{G}}\ket{\Psi_{\text{SD}}^N}\!.
  \label{eq:su3-nnhgroundstate}
\end{equation}
As $\ket{\Psi_{\text{SD}}^N}$ is an SU(3) singlet by construction and
$P_{\text{G}}$ commutes with SU(3) rotations, $\ket{\Psi_0}$ is an
SU(3) singlet as well.

If one interprets the state $\ket{0_\g}\equiv\prod_{\alpha=1}^N
c_{\alpha\g}^\dagger\vac$ as a reference state and the colorflip
operators $e^{\b\g}$ and $e^{\r\g}$ as ``particle creation
operators'', the state (\ref{eq:su3-nnhgroundstate}) can be rewritten
as~\cite{Kawakami92prb2,HaHaldane92}
\begin{equation}
  \ket{\Psi_0}=\sum_{\{z_i;w_k\}}\Psi_0[z_i;w_k]\;
  e_{z_1}^{\b\g}\ldots e_{z_{M_1}}^{\b\g}
  e_{w_1}^{\r\g}\ldots e_{w_{M_2}}^{\r\g}\ket{0_\g}\!,
\end{equation}
where the sum extends over all possible ways to distribute the
positions of the blue particles $z_1,\ldots,z_{M_1}$ and red particles
$w_1,\ldots,w_{M_2}$ over the $N$ sites.  The
vacuum state wave function is given by
\begin{equation}
    \Psi_0[z_i;w_k]\equiv
    \prod^{M_1}_{i<j}(z_i-z_j)^2\prod^{M_2}_{k<l}(w_k-w_l)^2
    \prod_{i=1}^{M_1}\prod_{k=1}^{M_2}(z_i-w_k)
    \prod_{i=1}^{M_1}z_i\prod_{k=1}^{M_2}w_k
  \label{eq:su3-definitionpsi0}
\end{equation}
with $M_1=M_2=M$, its energy is
\begin{equation}
  E_0=-\frac{\pi^2}{36}\left(N+\frac{15}{N}\right)\!.
  \label{eq:su3-gsenergyM=N/3}
\end{equation}
The total momentum, as defined through $e^{ip}=\Psi_0[\eta_1
z_i,\eta_1 w_k]/\Psi_0[z_i,w_k]$ with $\eta_1=\exp(i\frac{2\pi}{N})$,
is $p=0$ regardless of $M$. For further purposes, it is important to
note that the wave function \eqref{eq:su3-definitionpsi0} can be
equally expressed by any two sets of color variables, as it is shown in
App.~\ref{appsecrep}.

\subsection{Coloron excitations}\label{subsecsu3col}

Let $N=3M-1$, $M_1=(N-2)/3$, $M_2=(N+1)/3$. A localized coloron
at site "$\eta_\gamma$" is constructed by annihilation of a particle
with color $\sigma$ from a Slater determinant state of $N+1$ fermions
before Gutzwiller projection~\cite{Schuricht-06prb}:
\begin{equation}
\ket{\Psi_{\gamma \bar{\sigma}}^{\text{c}}}=P_{\text{G}}c_{\gamma \sigma}
\ket{\Psi_{\text{SD}}^{N+1}}, \label{onecol}
\end{equation}
where $\bar{\sigma}$ denotes the complementary color of the coloron.
The annihilation of the fermion causes an inhomogeneity in the SU(3)
spin and charge degree of freedom. The projection, however, smoothes
out the inhomogeneity in the charge degrees of freedom, the coloron
thus possesses color, but no charge.  The wave function of a
localized, \eg anti-blue or yellow, coloron is given by
\begin{equation}
\Psi_{\gamma}^{\text{c}}[z_i;w_k]=\prod_{i=1}^{M_1}(\eta_\gamma - z_i)
\Psi_0 [z_i;w_k], \label{wavecol}
\end{equation}
with $\Psi_0$ as stated in \eqref{eq:su3-definitionpsi0}. Fourier
transformation yields the momentum eigenstates
\begin{equation}
\label{onecolmom}
\ket{\Psi_n^{\text{c}}}=\frac{1}{N}\sum_{\gamma=1}^{N}(\bar{\eta}_\gamma)^n
\ket{\Psi_\gamma^{\text{c}}},
\end{equation}
which identically vanish unless $0 \leq n \leq M_1$. In particular,
this implies that the localized one-coloron states \eqref{onecol} form
an overcomplete set. It is hence not possible to interpret the
``coordinate'' $\eta_\gamma$ literally as the position of the coloron.
The momentum of \eqref{onecolmom} is
\begin{equation}
p^{\text{c}}_n=\frac{4\pi}{3}-\frac{2\pi}{N}\left(n+\frac{1}{3}\right),\ 
0 \leq n \leq M_1.
\end{equation}
The momentum eigenstates~\eqref{onecolmom} are found to be exact
energy eigenstates of the Hamiltonian~\eqref{ky1} with energies
\begin{equation}
E_n^{\text{c}}=E_0+\frac{2}{9}\frac{\pi^2}{N^2}+\epsilon^{\text{c}}
\bigl(p^{\text{c}}_n\bigr),
\end{equation}
where the one-coloron dispersion is given by
\begin{equation}
\epsilon^{\text{c}}(p)=\frac{3}{4}\left(\frac{\pi^2}{9}-(p-\pi)^2 \right).
\end{equation}
Colorons obey fractional statistics, the statistical parameter between
color-polarized colorons is given by $g=2/3$.

\subsection{One-Holon excitations}\label{subsecsu3one}

\subsubsection{One-holon wave functions}\label{2subsecsu3oneprop}

If we dope holes into the SU(3) spin chain, this will cause the
existence of holons, the elementary charge excitations of the system.
In this section, we will construct the wave functions of the one-holon
states and prove by explicit calculation that these states are
eigenstates of the Hamilitonian \eqref{ky1}. For this consider a chain
with $N=3M+1$ sites.  A localized holon at lattice site
$\eta_\xi$ is constructed as
\begin{equation}
\ket{\Psi_{\xi}^{\text{ho}}}=c_{\xi \sigma}^{\phantom{\dagger}}
P_{\text{G}}c_{\xi \sigma}^{\dagger}\ket{\Psi_{\text{SD}}^{N-1}},
\label{su3onehole}
\end{equation}
where the color index $\sigma$ can be chosen arbitrarily.  Compared to
the coloron, we eliminate the inhomogeneity in color while creating an
inhomogeneity in the charge distribution after Gutzwiller projection.
Thus the holon has no color but charge $e>0$ (as the charge at site
$\eta_\xi$ is removed).  Note that the holon is constructed as
apparently being strictly localized at the coordinate $\xi$, as states
\eqref{su3onehole} on neighboring coordinates are orthogonal. In
total, there are $N$ independent states of the form
\eqref{su3onehole}.

Momentum eigenstates are constructed from \eqref{su3onehole} by
Fourier transformation. We will show below that only $(N+5)/3 $ of
them are energy eigenstates, and restrict ourselves to this subset
in the following.  In order to describe these states by their wave
functions, we take $\ket{0_\text{g}}\equiv
\prod_{\alpha=1}^{N}c_{\alpha\text{g}}\ket{0}$ as reference state
and and write the one-holon states as
\begin{equation}
\ket{\Psi_m^{\text{ho}}}=\sum_{\{z_i;w_k;h\}}
\Psi_m^{\text{ho}}[z_i;w_k;h]\,c_{h\g}e_{z_1}^{\b\g}\dots e_{z_{M_1}}^{\b\g}
e_{w_1}^{\r\g}\dots e_{w_{M_2}}^{\r\g}\ket{0_{\g}}, 
\label{su3momonehole}
\end{equation}
where the sum extends over all possible ways to distribute the blue
coordinates $z_i$, the red coordinates $w_k$, and the holon coordinate
$h$ over the $N$ sites subject to the restriction $h \neq z_i,w_k$.
The one-holon wave function is given by
\begin{equation}
\Psi_{m}^{\text{ho}}[z_i;w_k;h]=h^{m}\prod_{i=1}^{M_{1}}(h-z_{i})
\prod_{k=1}^{M_{2}}(h-w_{k})\Psi_0[z_i;w_k].
\label{su3waveonehole}
\end{equation}
To increase readability of the following calculations, we will keep
the distinction between $M_1$ and $M_2$, although we will always set
$M_1$, $M_2$ and $M_3$, \ie the numbers of blue, red, and green
particles, to be equal to $M$ at the end.  In order for
\eqref{su3waveonehole} to represent energy eigenstates, the integer
$m$ has to be restricted to
\begin{equation}
0 \leq m \leq M+1=\frac{N+2}{3}. \label{su3restriction}
\end{equation}
For other values of $m$, the states $\ket{\Psi_m^{\text{ho}}}$ 
are not eigenstates of the Hamiltonian \eqref{ky1}, although they do
not vanish identically (as the $\ket{\Psi_n^{\text{c}}}$'s do).
Consequently, we are allowed to refer to the states
\eqref{su3momonehole} with \eqref{su3waveonehole} as ``holons'' only
if $0\leq m \leq M+1$.

This also implies that the states \eqref{su3onehole} do not really
constitute ``holons'' localized in position space, but only basis
states which can be used to construct holons if the momentum is chosen
adequately.  
Since the states \eqref{su3onehole} are orthogonal for different
lattice positions $\xi$, there are $N=3M+1$ orthogonal position basis
states $\ket{\Psi_\xi^{\text{ho}}}$.  These states cannot strictly be
holons, but rather constitute incoherent superpositions of holons and
other states. It is hence not possible to localize a holon onto a
single lattice site.  The best we can do is to take a Fourier
transform of the exact eigenstates $\ket{\Psi_m^{\text{ho}}}$ for
$0\leq m \leq M+1$ back into position space.  The resulting
``localized'' holon states will be true holons but will not be
localized strictly onto lattice sites.

The momentum of \eqref{su3momonehole} is
\begin{equation}
p_m^{\text{ho}}=\frac{2\pi}{3}+\frac{2\pi}{N}\left(m-\frac{1}{3} \right).
\label{momentumonesu3}
\end{equation}
The one-holon energies are derived below to be
\begin{equation}
E_m^{\text{ho}}=E_0-\frac{2}{9}\frac{\pi^2}{N^2}
+\epsilon^{\text{ho}}\bigl(p_m^{\text{ho}}\bigr), 
\label{su3energyonehole}
\end{equation}
where the one-holon dispersion is given by
\begin{equation}
\epsilon^{\text{ho}}(p)=
-\frac{3}{4}\left(\frac{\pi^2}{9}-(p-\pi)^2 \right), 
\quad \frac{2\pi}{3}\leq p \leq \frac{4\pi}{3}.
\label{su3oneholondispersion}
\end{equation}
In the following subsection we will prove that the states
\eqref{su3momonehole} are energy eigenstates of the Hamiltonian
\eqref{ky1}, if (and only if) the momentum quantum number $m$ is restricted to
\eqref{su3restriction}.

\subsubsection{Derivation of the one-holon energies}\label{2subsecsu3oneact}

To evaluate the action of $H_{\text{SU(3)}}$ on
$\ket{\Psi_m^{\text{ho}}}$, we first replace $e_{\alpha\!\!\!\phantom{\beta}}^{\g
  \g}e_{\beta}^{\g \g}$
by $(1-h_\alpha-e_\alpha^{\b\b}-e_\alpha^{\r\r})
(1-h_\beta-e_{\beta}^{\b\b}-e_{\beta}^{\r\r})$, where $h_\alpha$
denotes the hole occupation operator $h_\alpha=1-n_\alpha$, and
rewrite the Hamiltonian \eqref{kysu32} as
\begin{eqnarray}
  H_\text{SU(3)}&=&\phantom{+}\frac{2\pi^2}{N^2}
  \sum^N_{\alpha\neq\beta}\frac{1}{\vert\eta_\alpha-\eta_\beta\vert^2}
  \left(e_{\alpha\phantom{\!\!\!\beta}}^{\b\g}e_\beta^{\g\b}+
    e_{\alpha\phantom{\!\!\!\beta}}^{\r\g}e_\beta^{\g\r}+
    e_\alpha^{\b\r}e_\beta^{\r\b}\right)\nonumber\\
  & &+\frac{2\pi^2}{N^2}\sum^N_{\alpha\neq\beta}
  \frac{1}{\vert\eta_\alpha -\eta_\beta\vert^2}
  \left(e_\alpha^{\b\b}e_\beta^{\b\b}+
    e_\alpha^{\r\r}e_\beta^{\r\r}+e_\alpha^{\b\b}e_\beta^{\r\r}\right)
  \nonumber\\
  & &-\frac{2\pi^2}{N^2}\sum^N_{\alpha\neq\beta}
  \frac{1}{\vert\eta_\alpha -\eta_\beta\vert^2}
  \left(e_\alpha^{\b\b}+e_\alpha^{\r\r}\right)
  +\frac{2\pi^2}{N^2}\sum^N_{\alpha\neq\beta}
  \frac{1}{\vert\eta_\alpha -\eta_\beta\vert^2}
  \left(n_\alpha-\frac{1}{2}\right)\nonumber \\  
  & &+\frac{2\pi^2}{N^2}\sum^N_{\alpha\neq\beta}
  \frac{1}{\vert\eta_\alpha -\eta_\beta\vert^2}\big(e_\alpha^{\b\b}+
  e_\alpha^{\r\r}\big)(1-n_\beta)\nonumber \\
  & &+\frac{2\pi^2}{N^2} \sum_{\alpha \neq \beta}^N
  \frac{1}{\vert \eta_\alpha - \eta_\beta \vert^2}
  \Bigg[ \frac{1}{2}(c_{\alpha \b}^{\phantom{\dagger}}c_{\beta \b}^{\dagger}+
  c_{\alpha \r}^{\phantom{\dagger}}c_{\beta \r}^{\dagger})+
  \frac{1}{2}(c_{\alpha \b}^{\phantom{\dagger}}c_{\beta \b}^{\dagger}+
  c_{\alpha \g}^{\phantom{\dagger}}c_{\beta \g}^{\dagger})\nonumber \\
  & &\hspace{39mm}+\frac{1}{2}(c_{\alpha \r}^{\phantom{\dagger}}
  c_{\beta \r}^{\dagger}+c_{\alpha \g}^{\phantom{\dagger}}
  c_{\beta \g}^{\dagger}) \Bigg]. \label{hamiltonian} 
\end{eqnarray}
In the following we evaluate each term of \eqref{hamiltonian}
separately.

The first term
$\bigl[e_{\alpha\phantom{\!\!\!\beta}}^{\b\g}e_\beta^{\g\b}
\Psi_m^{\text{ho}}\bigr][z_i;w_k;h]$, which vanishes unless one of the
$z_i$'s is equal to $\eta_\alpha$, yields through Taylor expansion
(the derivative operators are understood to act on the analytic
extension of the wave function)
\begin{eqnarray}
&&\hspace{-15mm}\left[\sum_{\alpha\neq\beta}^N
\frac{e_\alpha^{\b\g}e_\beta^{\g\b}}{\vert\eta_\alpha-\eta_\beta\vert^2}
\Psi_m^{\text{ho}}\right]\!\![z_i;w_k;h]
=\sum_{i=1}^{M_1}\sum_{\beta\neq i}^N
\frac{\eta_\beta}{\vert z_i-\eta_\beta\vert^2}
\frac{\Psi_m^{\text{ho}}[\dots,z_{i-1},\eta_\beta,z_{i+1},\dots;w_k;h]}
{\eta_\beta}\nonumber\\
&=&\sum_{i=1}^{M_1}\sum_{\ell=0}^{N-1}
\frac{A_\ell z_i^{\ell+1}}{\ell!}
\frac{\partial^\ell}{\partial z_i^\ell}\frac{\Psi_m^{\text{ho}}}{z_i}
\label{eq:taylorexpansion}\\ 
&=&\frac{M_1}{12}\bigl(N^2+8M_1^2-6M_1(N+1)+3\bigr)\,
\Psi_m^{\text{ho}}\label{eq:1331}\\
& &-\frac{N-3}{2}\sum_{i=1}^{M_1}\sum_{k=1}^{M_2}
\frac{z_i}{z_i-w_k}\Psi_m^{\text{ho}}
+\sum_{i\neq j}^{M_1}\frac{z_i^2}{(z_i-z_j)^2}\Psi_m^{\text{ho}}\label{eq:1331a}\\ 
& &+2\sum_{i\neq j}^{M_1}\sum_{k=1}^{M_2}
\frac{z_i^2}{(z_i-z_j)(z_i-w_k)}\Psi_m^{\text{ho}}\label{eq:1331b}\\
& &+\frac{1}{2}\sum_{i=1}^{M_1}\sum_{k\neq l}^{M_2}
\frac{z_i^2}{(z_i-w_k)(z_i-w_l)}\Psi_m^{\text{ho}}\label{eq:1331c}\\
& &+\sum_{i\neq j}^{M_1}\frac{2z_i^2}{(z_i-z_j)(z_i-h)}
\Psi_{m}^{\ho}-\frac{N-3}{2}\sum_{i=1}^{M_1}\frac{z_i}{z_i-h}\Psi_{m}^{\ho} 
\label{eq:appsu3-5} \\
& &+\sum_{i=1}^{M_1}\sum_{k=1}^{M_2}\frac{z_i^2}{(z_i-w_k)(z_i-h)}
\Psi_{m}^{\ho}\; , \label{eq:appsu3-6} 
\end{eqnarray}
where we have used $\text{deg}_{z_i}\Psi_m^{\text{ho}}[z_i;w_k;h]=N-1$
and defined $A_\ell\equiv-\sum_{\alpha=1}^{N-1} \eta_\alpha^2
(\eta_\alpha -1)^{\ell-2}$.  Evaluation of the latter yields
$A_0=(N-1)(N-5)/12$, $A_1=-(N-3)/2$, $A_2=1$, and $A_\ell=0$ for
$2<\ell\le N-1$ (see App.~\ref{app:bseries}).
Furthermore, we have used
\begin{equation}
\frac{x^2}{(x-y)(x-z)}+\frac{y^2}{(y-x)(y-z)}+\frac{z^2}{(z-x)(z-y)}=1,
\quad x,y,z\in\mathbb{C}.
\label{eq:appsu3-threezformula}
\end{equation}

The second term
$\bigl[e_{\alpha\phantom{\!\!\!\beta}}^{\r\g}e_\beta^{\g\r}
\Psi_m^{\text{ho}}\bigr][z_i;w_k;h]$ can be treated in the same way,
yielding together with the first term in \eqref{eq:1331a}
\begin{displaymath}
-\frac{N-3}{2}\sum_{i=1}^{M_1}\sum_{k=1}^{M_2}\frac{z_i}{z_i-w_k}
+\frac{N-3}{2}\sum_{i=1}^{M_1}\sum_{k=1}^{M_2}\frac{w_k}{z_i-w_k}=
-\frac{N-3}{2}M_1M_2.
\end{displaymath} 
One part of (\ref{eq:1331b}) and the term corresponding to
(\ref{eq:1331c}) can be simplified with
\eqref{eq:appsu3-threezformula} to
\begin{displaymath}
\sum_{i\neq j}^{M_1}\sum_{k=1}^{M_2}\left(
\frac{z_i^2}{(z_i-z_j)(z_i-w_k)}+\frac{1}{2}\frac{w_k^2}{(z_i-w_k)(z_j-w_k)}
\right)=\frac{1}{2}M_1(M_1-1)M_2,
\end{displaymath}
as well as similar expressions for $z_i\leftrightarrow w_k$.

The third term $\bigl[e_{\alpha\phantom{\!\!\!\beta}}^{\b\r}e_\beta^{\r\b}
\Psi_m^{\text{ho}}\bigr][z_i;w_k;h]$ leads to
\begin{eqnarray}
&&\hspace{-10mm}\left[\sum_{\alpha\neq\beta}^N
\frac{e_\alpha^{\b\r}e_\beta^{\r\b}}{\vert\eta_\alpha-\eta_\beta\vert^2}
\Psi_m^{\text{ho}}\right]\!\![z_i;w_k;h]\nonumber\\
&=&\sum_{i=1}^{M_1}\sum_{k=1}^{M_2}\frac{z_iw_k}{(z_i-w_k)^2}
\prod^{M_1}_{j\neq i}\left(1+\frac{z_i-w_k}{z_j-z_i}\right)
\prod^{M_2}_{l\neq k}\left(1-\frac{z_i-w_k}{w_l-w_k}\right)
\Psi_m^{\text{ho}}[z_i;w_k;h]\nonumber\\
&=&\phantom{-}\sum_{i=1}^{M_1}\sum_{k=1}^{M_2}
\frac{z_iw_k}{(z_i-w_k)^2}\Psi_m^{\text{ho}}\label{eq-appsu3-1221first}\\
& &-\sum_{i\neq j}^{M_1}\sum_{k=1}^{M_2}
\frac{z_iw_k}{(z_i-z_j)(z_i-w_k)}\Psi_m^{\text{ho}}-
\sum_{i=1}^{M_1}\sum_{k\neq l}^{M_2}
\frac{z_iw_k}{(w_k-z_i)(w_k-w_l)}\Psi_m^{\text{ho}}
\label{eq:appsu3-1221second}\\
& &+\sum_{i=1}^{M_1}\sum_{k=1}^{M_2}\sum_{\mu=2}^{M_1-1}\frac{1}{\mu!}
\sum_{\{a_j\}}\frac{z_iw_k(z_i-w_k)^{\mu-2}}{(z_{a_1}-z_i)\cdots
(z_{a_\mu}-z_i)}\Psi_m^{\text{ho}}\label{eq:appsu3-1221third}\\
& &+\sum_{i=1}^{M_1}\sum_{k=1}^{M_2}\sum_{\nu=2}^{M_2-1}\frac{(-1)^\nu}{\nu!}
\sum_{\{b_l\}}\frac{z_iw_k(z_i-w_k)^{\nu-2}}{(w_{b_1}-w_k)\cdots
(w_{b_\nu}-w_k)}\Psi_m^{\text{ho}}\label{eq:appsu3-1221forth}\\
& &+\sum_{i=1}^{M_1}\sum_{k=1}^{M_2}\sum_{\mu=1}^{M_1-1}\sum_{\nu=1}^{M_2-1}
\frac{(-1)^\nu}{\mu!\nu!}\nonumber\\
& &\quad\times\sum_{\{a_j;b_l\}}
\frac{z_iw_k(z_i-w_k)^{\mu+\nu-2}}{(z_{a_1}-z_i)\cdots(z_{a_\mu}-z_i)
(w_{b_1}-w_k)\cdots(w_{b_\nu}-w_k)}\Psi_m^{\text{ho}}
\label{eq:appsu3-1221fifth},
\end{eqnarray}
where $\{a_j\}$ ($\{b_l\}$) is a set of integers between 1 and $M_1$ ($M_2$).
The summations run over all possible ways to distribute the $z_{a_j}$
($w_{b_l}$) over the blue (red) coordinates, where $z_i$ ($w_k$) is excluded.
The two terms (\ref{eq:appsu3-1221third}) and (\ref{eq:appsu3-1221forth}) 
vanish due to~\cite{Schuricht-06prb}
\begin{theo}\label{theo:gstheorem}
  Let $M\ge 3$, $z\in\mathbb{C}$, and $z_1,\ldots, z_M\in\mathbb{C}$ distinct.
  Then,
\begin{equation}
\sum_{i=1}^M\frac{z_i(z_i-z)^{M-3}}{\prod_{j\neq i}^M(z_j-z_i)}=0.
\label{eq:theo1}
\end{equation}
\end{theo}
The last term (\ref{eq:appsu3-1221fifth}) can be simplified using a
theorem due to Ha and Haldane~\cite{HaHaldane92}:
\begin{theo}\label{theo:appsu3-hahaldane}
  Let $\{a_j\}$ be a set of distinct integers between $1$ and $M_1$, and
  $\{b_l\}$ a set of distinct integers between $1$ and $M_2$. Then,
\begin{displaymath}
\begin{split}
\sum_{i=1}^{M_1}\sum_{k=1}^{M_2}\sum_{\mu=1}^{M_1-1}\sum_{\nu=1}^{M_2-1}
\sum_{\{a_j;b_l\}}\frac{(-1)^\nu}{\mu!\nu!}&
\frac{z_iw_k(z_i-w_k)^{\mu+\nu-2}}{(z_{a_1}-z_i)\cdots(z_{a_\mu}-z_i)
(w_{b_1}-w_k)\cdots(w_{b_\nu}-w_k)}\\
&=-\sum_{\kappa=1}^{\mathrm{min}(M_1,M_2)}\!\!\!\!(M_1-\kappa)(M_2-\kappa).
\end{split}
\end{displaymath}
\end{theo}
Furthermore, the two terms in line (\ref{eq:appsu3-1221second}),
together with the remainder of (\ref{eq:1331b}) and the corresponding
expression from the second term of the Hamiltonian, can be simplified
to $M_1M_2(M_1+M_2-2)\Psi_m^{\text{ho}}/2$. 

The 2nd and 3rd line of \eqref{hamiltonian} yield
\begin{eqnarray}
& &\hspace{-8mm}\sum_{\alpha \neq \beta}^{N}
\frac{e_{\alpha}^{\bb}e_{\beta}^{\bb}+e_{\alpha}^{\rr}e_{\beta}^{\rr}+
e_{\alpha}^{\bb}e_{\beta}^{\rr}-e_{\alpha}^{\b\b}-e_{\alpha}^{\r\r}+n_\alpha-
\frac{1}{2}}{|\eta_{\alpha}-\eta_{\beta}|^2}\Psi_{m}^{\ho}[z_i;w_k;h]
\nonumber \\
&=&\frac{1}{2}\bigl(M_1(M_1-1)+M_2(M_2-1)\bigr)
\Psi_{m}^{\text{ho}}-\sum_{i\neq j}^{M_1}
\frac{z_{i}^2}{(z_i-z_j)^2}\Psi_{m}^{\text{ho}}-
\sum_{k \neq l}^{M_2}
\frac{w_{k}^{2}}{(w_k - w_l)^2}\Psi_{m}^{\text{ho}}
\nonumber \\*
& &-\sum_{i=1}^{M_1}\sum_{k=1}^{M_2}\frac{z_iw_k}{(z_i-w_k)^2}
\Psi_{m}^{\text{ho}}-\frac{N^2-1}{12}\left(M_1+M_2-\frac{N}{2}+1\right) 
\Psi_{m}^{\ho}, 
\label{su3j1}
\end{eqnarray}
by which the remainder of \eqref{eq:1331a}, its counterpart from the
second term, and \eqref{eq-appsu3-1221first} are cancelled.

The 4th line of \eqref{hamiltonian} yields
\begin{eqnarray}
& &\hspace{-8mm}\sum_{\alpha \neq \beta}^{N}
\frac{\big(e_\alpha^{\b\b}+e_\alpha^{\r\r}\big)(1-n_\beta)}
{|\eta_{\alpha}-\eta_{\beta}|^2}\Psi_{m}^{\ho}[z_i;w_k;h]=
\left(\sum_{i=1}^{M_1}\frac{1}{\vert z_i - h \vert^2}+
\sum_{k=1}^{M_2}\frac{1}{\vert w_k-h \vert^2}\right)\Psi_m^{\text{ho}}. 
\label{absspin1}
\end{eqnarray}

We will now evaluate the charge kinetic terms, which include the
technical improvements compared to previous calculations.  We will use
a Taylor expansion as in \eqref{eq:taylorexpansion}.
For the treatment of the charge kinetic terms it is crucial that the
fermionic creation and annihilation operators appearing in the
expansion match with the variables of the analytically extended wave
function.  We wish to stress that the one-holon wave function
\eqref{su3waveonehole} can, as the ground state wave function, be
equally expressed by an arbitrary pair of sets of color
variables. 
In \eqref{hamiltonian} we have thus written the charge kinetic terms
in a symmetricized way.  For the first term we get
\begin{eqnarray}
&&\hspace{-10mm}\left[ \sum_{\alpha\neq\beta}^N \frac{1}{2}
\frac{c_{\alpha \b}^{\phantom{\dagger}}c_{\beta \b}^{\dagger}+
c_{\alpha \r}^{\phantom{\dagger}}c_{\beta \r}^{\dagger}}
{\vert \eta_\alpha - \eta_\beta \vert^2}\Psi_{m}^{\ho}\right]\!\![z_i;w_k;h]=
\left[\sum_{\alpha \neq \beta}^N
\frac{c_{\alpha v}^{\phantom{\dagger}}c_{\beta v}^{\dagger}}
{\vert \eta_\alpha - \eta_\beta \vert^2}\Psi_{m}^{\ho}\right]\!\!
[v_{1},\dots,v_{M_1+M_2};h]\nonumber \\
&=&\sum_{\beta \neq h}^N
\frac{\eta_{\beta}^{m}}{\vert h-\eta_\beta\vert^2}
\frac{\Psi_{m}^{\ho} [v_{1},\dots,v_{M_1+M_2};\eta_{\beta}]}
{\eta_\beta^{m}}\nonumber \\
&=& \sum_{\ell = 0}^{M_1+M_2}
\sum_{\beta \neq h}^N
\frac{\eta_\beta^{m}(\eta_\beta - h)^\ell}
{\ell!\vert h - \eta_\beta\vert^2}
\frac{\partial^\ell}{\partial \eta_{\beta}^\ell}
\left( \frac{ \Psi_{m}^{\ho}[v_{1},\dots,v_{M_1+M_2};\eta_{\beta}]}
{\eta_{\beta}^{m}} \right)\rule[-16pt]{0.3pt}{26pt}_{\,\eta_{\beta}=h} 
\nonumber \\
&=&\sum_{\ell = 0}^{M_1+M_2}\frac{B_\ell^mh^{m+\ell}}{\ell!}
\frac{\partial^\ell}{\partial h^\ell}
\left(\frac{\Psi_{m}^{\ho}[z_{1},\dots,z_{M_1};w_{1},\dots,w_{M_2};h]}
{h^{m}} \right) \nonumber \\
&=&\left[\biggl(\frac{N^2-1}{12}+\frac{m(m-N)}{2}\biggr)h^{m}-
\biggl(\frac{N-1}{2}-m\biggr)h^{m+1}\frac{\partial}{\partial h}
+\frac{1}{2} h^{m+2}\frac{\partial^{2}}{\partial h^{2}} 
\right] \frac{\Psi_{m}^{\ho}[z_{i};w_{k};h]}{h^{m}}\nonumber \\
&=& \biggl(\frac{N^2-1}{12}+\frac{m(m-N)}{2}\biggr)\Psi_m^{\text{ho}} -
\biggl(\frac{N-1}{2}-m\biggr)\left[\sum_{i=1}^{M_1}\frac{h}{h-z_i}+
\sum_{k=1}^{M_2}\frac{h}{h-w_k}\right]\Psi_m^{\text{ho}} \nonumber\\*
& &\!\!\!+\frac{1}{2}\left[\sum_{i \neq j}^{M_1}
\frac{h^2}{(h-z_i)(h-z_j)} +\sum_{i=1}^{M_1}\sum_{k=1}^{M_2}
\frac{2h^2}{(h-z_i)(h-w_k)}+\sum_{k \neq l}^{M_2}\frac{h^2}{(h-w_k)(h-w_l )} 
\right]\Psi_m^{\text{ho}}, \label{su3oneholoncont1}
\end{eqnarray}
where the $v_i$'s denote the union of the blue and the red
coordinates, and we have introduced
$B_\ell^m=-\sum_{\alpha=1}^{N-1}\eta_\alpha^{m+1}(\eta_\alpha-1)^{\ell-2}$.
Evaluation of the latter yields $B_0^m=(N^2-1)/12+m(m-N)/2$,
$B_1^m=m-(N-1)/2$, $B_2^m=1$, and $B_\ell=0$ for $3 \leq \ell$ and
$0\leq m \leq (N+2)/3$ (see App.~\ref{app:bseries}).  
The restriction \eqref{su3restriction} of the allowed momentum values
follows from the $B$-series in \eqref{su3oneholoncont1}, since
$B_\ell\neq 0$ for $3 \leq \ell$ and $(N+2)/3<m$, in which case the
calculations above are not feasible anymore.

For the evaluation of the remaining two charge kinetic terms we
reexpress the wave function $\Psi_m^{\text{ho}}$ by the other pairs of
sets of color variables (see App.~\ref{appsecrep}).  We then proceed
as in \eqref{su3oneholoncont1}, where we replace the green variables
by the blue and red ones using the identities of
App.~\ref{appsecderive}.  Doing so, we finally arrive at
\begin{eqnarray}
&&\hspace{-10mm}\left[\sum_{\alpha\neq \beta}^N \frac{1}{2}
\frac{c_{\alpha \b}^{\phantom{\dagger}}c_{\beta \b}^{\dagger}+
c_{\alpha \g}^{\phantom{\dagger}}c_{\beta \g}^{\dagger}}{\vert \eta_\alpha - 
\eta_\beta \vert^2}\Psi_{m}^{\ho}\right]\!\![z_i;w_k;h]\nonumber \\
&=&\biggl(\frac{N^2-1}{12}+\frac{m(m-N)}{2}\biggr)\Psi_{m}^{\ho}-
\biggl(\frac{N-1}{2}-m\biggr)\Bigg[C_1-\sum_{k=1}^{M_2}\frac{h}{h-w_k} 
\Bigg]\Psi_{m}^{\ho}\nonumber \\*
& &+\frac{1}{2}\Bigg[C_1^2-C_2 -2C_1\sum_{k=1}^{M_2}
\frac{h}{h-w_k}+2\sum_{k=1}^{M_2}\frac{h^2}{(h-w_k)^2}+
\sum_{k \neq l}^{M_2}\frac{h^2}{(h-w_k)(h-w_l )} 
\Bigg]\Psi_{m}^{\ho}, 
\label{su3chargebg1}
\end{eqnarray}
as well as
\begin{eqnarray}
&&\hspace{-10mm}\left[\sum_{\alpha\neq \beta}^N \frac{1}{2}
\frac{c_{\alpha \r}^{\phantom{\dagger}}c_{\beta \r}^{\dagger}+
c_{\alpha \g}^{\phantom{\dagger}}c_{\beta \g}^{\dagger}}{\vert \eta_\alpha - 
\eta_\beta \vert^2}\Psi_{m}^{\ho}\right]\!\![z_i;w_k;h]\nonumber \\
&=&\biggl(\frac{N^2-1}{12}+\frac{m(m-N)}{2}\biggr)\Psi_{m}^{\ho}-
\biggl(\frac{N-1}{2}-m\biggr)\Bigg[C_1-\sum_{i=1}^{M_1}\frac{h}{h-z_i} 
\Bigg]\Psi_{m}^{\ho}\nonumber \\*
& &+\frac{1}{2}\Bigg[C_1^2-C_2 -2C_1\sum_{i=1}^{M_1}\frac{h}{h-z_i}+2
\sum_{i=1}^{M_1}\frac{h^2}{(h-z_i)^2}+\sum_{i \neq j}^{M_1}
\frac{h^2}{(h-z_i)(h-z_j)} \Bigg] \Psi_{m}^{\ho}. 
\label{su3chargerg2}
\end{eqnarray}
In \eqref{su3chargebg1} and \eqref{su3chargerg2} we have defined the
constants $C_1=\sum_{\alpha=1}^{N-1}1/(1-\eta_{\alpha})=(N-1)/2$ and
$C_2=\sum_{\alpha=1}^{N-1}1/(1-\eta_{\alpha})^{2}=-(N^2-6N+5)/12$ (see
App.~\ref{app:bseries}). 

Now there occur several simplifications. The first term in
\eqref{eq:appsu3-5} together with the appropriate terms in
\eqref{su3oneholoncont1} and \eqref{su3chargerg2} yield
\begin{equation}
\sum_{i \neq j}^{M_1}\frac{2z_{i}^{2}}{(z_i-z_j)(z_i-h)}+
\sum_{i \neq j}^{M_1}\frac{h^2}{(h-z_i)(h-z_j)}=M_1(M_1-1),
\end{equation}
and the similar expression for $z_i \leftrightarrow w_k$ leads to
$M_2(M_2-1)$. Furthermore, \eqref{eq:appsu3-6}, its counterpart from
the second term, and the appropriate term in \eqref{su3oneholoncont1}
result in
\begin{equation}
\sum_{i=1}^{M_1}\sum_{k=1}^{M_2}\frac{z_{i}^{2}}{(z_i-w_k)(z_i-h)}
+\sum_{i=1}^{M_1}\sum_{k=1}^{M_2}\frac{w_k^{2}}{(w_k-z_i)(w_k-h)} +
\sum_{i=1}^{M_1}\sum_{k=1}^{M_2}\frac{h^2}{(h-z_i)(h-w_k)}=M_1 M_2.
\end{equation}
Finally, the remainder of \eqref{eq:appsu3-5}, the first term of
\eqref{absspin1}, and the remaining off-diagonal terms of
\eqref{su3chargerg2} yield
\begin{equation}
-\frac{N-3}{2}\sum_{i=1}^{M_1}\frac{z_i}{z_i-h}+
\sum_{i=1}^{M_1}\frac{1}{|z_i-h|^2}-
\frac{N-1}{2}\sum_{i=1}^{M_1}\frac{h}{h-z_i}+
\sum_{i=1}^{M_1}\frac{h^2}{(h-z_i)^2}=-M_1\frac{N-3}{2},
\end{equation}
and the similar expression for $z_i \leftrightarrow w_k$.

Summing up all terms, we obtain
\begin{equation}
H_{\text{SU(3)}}\ket{\Psi_m^{\text{ho}}}=E_m^{\text{ho}}\ket{\Psi_m^{\text{ho}}}
\end{equation}
with
\begin{equation}
E_m=-\frac{2\pi^2}{N^2}\left[\frac{1}{72}N^3+\frac{1}{24}N-
\frac{1}{18}-\frac{3}{2}m\left(m-\frac{N+2}{3}\right)\right],
\end{equation}
where we have set $M_1=M_2=M=(N-1)/3$. Using \eqref{momentumonesu3},
$E_m$ can now be easily brought into the form \eqref{su3energyonehole}.

\subsection{Two-Holon excitations}\label{subsecsu3two}

\subsubsection{Momentum eigenstates}\label{2subsecsu3twoprop}

We will now investigate the two-holon eigenstates. For this, let the
number of sites be given by $N=3M+2$. The state with two localized
holons 
is constructed as
\begin{equation}
\ket{\Psi_{\xi_1 \xi_2}^{\text{ho}}}=c_{\xi_1 \sigma}^{\phantom{\dagger}}
c_{\xi_2 \sigma}^{\phantom{\dagger}}P_{\text{G}}c_{\xi_1 \sigma}^{\dagger}
c_{\xi_2 \sigma}^{\dagger}\ket{\Psi^{N-2}_{\text{SD}}}.
\label{su3twoholes}
\end{equation}
Similar to the one-holon case, these localized states
\eqref{su3twoholes} do not really represent ``holons'' localized in
position space, and we can refer to true physical holons only in
momentum space. The two-holon momentum eigenstates will be most easily 
described by their wave functions
\begin{equation}
\begin{split}
\Psi_{mn}^{\text{ho}}[z_i;w_k;h_1,h_2]=&(h_1-h_2)(h_1^m h_2^n +h_1^n h_2^m)\\
&\times\prod_{i=1}^{M_1}(h_1-z_i)(h_2-z_i)\prod_{k=1}^{M_2}(h_1-w_k)(h_2-w_k)
\Psi_0[z_i;w_k],
\label{su3wavetwoholes}
\end{split}
\end{equation} 
where $h_{1,2}$ denote the holon coordinates and the integers $m$ and
$n$ are restricted to
\begin{equation}
0 \leq n \leq m \leq M+1=\frac{N+1}{3}.
\label{su3tworestriction}
\end{equation} 
This restriction will be derived below.

The two-holon state represented by \eqref{su3wavetwoholes} is
\begin{equation}
\label{su3momtwoholes}
\ket{\Psi_{mn}^{\text{ho}}}=\sum_{\{z_i;w_k;h_1,h_2\}}
\Psi_{mn}^{\text{ho}}[z_i;w_k;h_1,h_2]\,c^{\phantom{\dagger}}_{h_1 \g}
c^{\phantom{\dagger}}_{h_2 \g}e_{z_1}^{\b\g}\dots e_{z_{M_1}}^{\b\g}
e_{w_1}^{\r\g}\dots e_{w_{M_2}}^{\r\g}\ket{0_{\g}},
\end{equation}
where the sum contains the restriction $h_{1,2} \neq z_i,w_k$.  The
total momentum of the states \eqref{su3momtwoholes} is found to be
\begin{equation}
p_{mn}^{\text{ho}}=\frac{4\pi}{3}+\frac{2\pi}{N}\left(m+n-\frac{1}{3} 
\right) \; \text{mod }2\pi. \label{momentumtwoholes} 
\end{equation}

In the following two subsections we construct the two-holon energy
eigenstates starting from \eqref{su3twoholes}. The used strategy is
similar to the construction of the two-holon states in the SU(2)
KYM~\cite{TSG}.

\subsubsection{Action of $H_{\text{SU(3)}}$ on the momentum eigenstates}
\label{2subsecsu3twoact}

In order to derive the action of the Hamiltonian on the momentum
eigenstates \eqref{su3momtwoholes}, we first define the auxiliary wave
functions
\begin{eqnarray}
\varphi_{mn}[z_i;w_k;h_1,h_2]&=&h_1^m h_2^n 
\prod_{i=1}^{M_1}(h_1-z_i)(h_2-z_i)
\prod_{k=1}^{M_2}(h_1-w_k)(h_2-w_k)\Psi_0[z_i;w_k] \nonumber \\
&\equiv&\psi_{h_1 h_2}\Psi_0[z_i;w_k],
\end{eqnarray}
which can be used to express the wave functions
\eqref{su3wavetwoholes} as 
\begin{equation}
\Psi_{mn}^{\text{ho}}=\varphi_{m+1,n}+\varphi_{n+1,m}-\varphi_{m,n+1}-
\varphi_{n,m+1}.
\label{phidecomposition}
\end{equation}

In agreement to the one-holon case, we use \eqref{hamiltonian} for the
Hamiltonian and concentrate on the terms which differ from the ones
above.  The first term
$\bigl[e_{\alpha\!\!\phantom{\beta}}^{\b\g}e_\beta^{\g\b}
\varphi_{mn}\bigr][z_i;w_k;h_1,h_2]$ yields
\begin{eqnarray}
&&\hspace{-15mm}\left[\sum_{\alpha\neq\beta}^N
\frac{e_\alpha^{\b\g}e_\beta^{\g\b}}{\vert\eta_\alpha-\eta_\beta\vert^2}
\varphi_{mn}\right]\!\![z_i;w_k;h_1,h_2]
=\sum_{i=1}^{M_1}\sum_{\ell=0}^{N-1}
\frac{A_\ell z_i^{\ell+1}}{\ell!}
\frac{\partial^\ell}{\partial z_i^\ell}\frac{\varphi_{mn}}{z_i}\nonumber\\ 
&=&\frac{M_1}{12}(N^2+8M_1^2-6M_1(N+1)+3)\,\varphi_{mn}\nonumber\\
& &-\frac{N-3}{2}\sum_{i=1}^{M_1}\sum_{k=1}^{M_2}
\frac{z_i}{z_i-w_k}\varphi_{mn}
+\sum_{i\neq j}^{M_1}\frac{z_i^2}{(z_i-z_j)^2}\varphi_{mn}\label{twosu3bg4}\\ 
& &+2\sum_{i\neq j}^{M_1}\sum_{k=1}^{M_2}
\frac{z_i^2}{(z_i-z_j)(z_i-w_k)}\Psi_m^{\text{ho}}+\frac{1}{2}
\sum_{i=1}^{M_1}\sum_{k\neq l}^{M_2}
\frac{z_i^2}{(z_i-w_k)(z_i-w_l)}\varphi_{mn}\label{twosu3bg3}\\
& &+\Psi_0\sum_{i=1}^{M_1}\left(\frac{1}{2}z_i^2
\frac{\partial^2}{\partial z_i^2}+\sum_{j\neq i}^{M_1}
\frac{2z_i^2}{z_i-z_j}\frac{\partial}{\partial z_i}-\frac{N-3}{2}z_i
\frac{\partial}{\partial z_i} \right)\psi_{h_1h_2} \label{twosu3bg1} \\
& &+\Psi_0\sum_{i=1}^{M_1}\sum_{k=1}^{M_2}\frac{z_i^2}{z_i-w_k}
\frac{\partial}{\partial z_i}\psi_{h_1h_2}. \label{twosu3bg2}
\end{eqnarray}
Now, the lines \eqref{twosu3bg4} and \eqref{twosu3bg3} can be treated
as in the one-holon calculation, whereas the lines \eqref{twosu3bg1}
and \eqref{twosu3bg2} explicitly yield using
\eqref{eq:appsu3-threezformula}:
\begin{eqnarray}
&&\phantom{+}\sum_{i=1}^{M_1}\left(1-\frac{h_1^2}{(h_1-h_2)(h_1-z_i)}-
\frac{h_2^2}{(h_2-h_1)(h_2-z_i)}\right)\varphi_{mn}\nonumber\\
& &+\sum_{i\neq j}^{M_1}\left(1-\frac{h_1^2}{(h_1-z_i)(h_1-z_j)}\right)
\varphi_{mn}+\sum_{i\neq j}^{M_1}\left(1-
\frac{h_2^2}{(h_2-z_i)(h_2-z_j)}\right)\varphi_{mn}\nonumber\\
& &-\frac{N-3}{2}\sum_i^{M_1}\left(1-\frac{h_1}{h_1-z_i}\right)\varphi_{mn}-
\frac{N-3}{2}\sum_i^{M_1}\left(1-\frac{h_2}{h_2-z_i}\right)\varphi_{mn}
\nonumber\\
& &+\sum_{i=1}^{M_1}\sum_{k=1}^{M_2}\left(1-\frac{w_k^2}{(w_k-z_i)(w_k-h_1)}-
\frac{h_1^2}{(h_1-z_i)(h_1-w_k)}\right)\varphi_{mn}\nonumber\\
& &+\sum_{i=1}^{M_1}\sum_{k=1}^{M_2}\left(1-\frac{w_k^2}{(w_k-z_i)(w_k-h_2)}-
\frac{h_2^2}{(h_2-z_i)(h_2-w_k)}\right)\varphi_{mn}. 
\end{eqnarray}
The term $\bigl[e_{\alpha\!\!\phantom{\beta}}^{\r\g}e_\beta^{\g\r}
\varphi_{mn}\bigr][z_i;w_k;h_1,h_2]$ leads to the analog result with
$z_i$ and $w_k$ interchanged. Furthermore, the terms
$\bigl[e_{\alpha\!\!\phantom{\beta}}^{\b\r}e_\beta^{\r\b}
\varphi_{mn}\bigr][z_i;w_k;h_1,h_2]$ as well as the 2nd and 3rd line
of \eqref{hamiltonian} are unchanged as compared to the one-holon
case. The 4th line of \eqref{hamiltonian} yields
\begin{equation}
\sum_{\alpha \neq \beta}^{N}
\frac{e_\alpha^{\b\b}+e_\alpha^{\r\r}}{|\eta_{\alpha}-
\eta_{\beta}|^2}(1-n_\beta)\,\varphi_{mn}=\left(\left\{\sum_{i=1}^{M_1}
\frac{1}{\vert z_i - h_1 \vert^2}+\sum_{k=1}^{M_2}
\frac{1}{\vert w_k-h_2\vert^2}\right\}
+\{h_1\leftrightarrow h_2\}\right)\varphi_{mn}, 
\label{twoholespinabs}
\end{equation}
where $\{h_1 \leftrightarrow h_2\}$ denotes the reappearance of the
preceeding terms in curly brackets with $h_1$ and $h_2$ interchanged.

For the charge kinetic terms we obtain in analogy to the one-holon
states
 \begin{eqnarray}
&&\hspace{-8mm}\left[ \sum_{\alpha\neq\beta}^N \frac{1}{2}
\frac{c_{\alpha \b}c_{\beta \b}^{\dagger}+c_{\alpha \r}c_{\beta \r}^{\dagger}}
{\vert \eta_\alpha - \eta_\beta \vert^2}\varphi_{mn}\right]\!\!
[z_i;w_k;h_1,h_2]=\left[\sum_{\alpha=h_1,h_2}\sum_{\beta\neq\alpha}^N
\frac{c_{\alpha v}^{\phantom{\dagger}}c_{\beta v}^{\dagger}}{\vert \eta_\alpha - 
\eta_\beta \vert^2}\varphi_{mn}\right]\!\![v_i;h_1,h_2]
\nonumber \\
&=&\sum_{\ell = 0}^{M_1+M_2}\frac{B_\ell^mh_1^{m+\ell}}{\ell!}
\frac{\partial^\ell}{\partial h_{1}^\ell}
\left( \frac{\varphi_{mn} }{h_1^{m}} \right)+\sum_{\ell = 0}^{M_1+M_2}
\frac{B_\ell^nh_2^{n+\ell}}{\ell!}\frac{\partial^\ell}{\partial h_{2}^\ell}
\left( \frac{\varphi_{mn} }{h_2^{n}} \right) \nonumber \\
&=&\Bigg[\left(\frac{N^2-1}{6}+\frac{m(m-N)}{2}+\frac{n(n-N)}{2}\right)
h_1^mh_2^n-\left(\frac{N-1}{2}-m\right)h_1^{m+1}h_2^n
\frac{\partial}{\partial h_1}\nonumber\\
& &-\left(\frac{N-1}{2}-n\right)h_1^{m}h_2^{n+1}\frac{\partial}{\partial h_2}
+\frac{1}{2}h_1^{m+2}h_2^n\frac{\partial^{2}}{\partial h_1^{2}}
+\frac{1}{2}h_1^mh_2^{n+2}\frac{\partial^2}{\partial h_2^2} \Bigg] 
\frac{\varphi_{mn}}{h_1^{m}h_2^n} \nonumber \\
&=&\Bigg[\frac{N^2-1}{6}+\frac{m(m-N)}{2}+\frac{n(n-N)}{2}
-\left(\frac{N-1}{2}-m  \right)\left(\sum_{i=1}^{M_1}\frac{h_2}{h_2-z_i}+
\sum_{k=1}^{M_2}\frac{h_2}{h_2-w_k} \right)\nonumber\\
& &-\left(\frac{N-1}{2}-n\right)\left(\sum_{i=1}^{M_1}\frac{h_1}{h_1-z_i}+
\sum_{k=1}^{M_2}\frac{h_1}{h_1-w_k} \right)\nonumber\\
& &+\Biggl\{\frac{1}{2}\sum_{i \neq j}^{M_1}\frac{h_1^2}{(h_1-z_i)(h_1-z_j)}+
\sum_{i=1}^{M_1}\sum_{k=1}^{M_2}\frac{h_1^2}{(h_1-z_i)(h_1-w_k)}\nonumber\\
& &\hspace{5mm}+\frac{1}{2}\sum_{k \neq l}^{M_2}\frac{h_1^2}{(h_1-w_k)(h_1-w_l )} 
\Biggr\}+\{h_1 \leftrightarrow h_2\} \Bigg]\varphi_{mn}.
\end{eqnarray}
In this term, the restriction of the allowed momentum eigenvalues
\eqref{su3tworestriction} follows from the $B$-series as in the
one-holon case.

The other charge kinetic terms are treated by using the fact that the
two-holon wave function can be expressed by either pairs of color
variables, as is shown in App.~\ref{appsecrep}. The terms involving
green variables are rewritten in terms of the $z_i$'s and $w_k$'s by
the identities given in App.~\ref{appsecderive}. Thus we finally
deduce for the sum of the three charge kinetic terms
\begin{eqnarray}
& &\hspace{-10mm}\left[\sum_{\alpha\neq\beta}^N\sum_{\sigma}
\frac{c_{\sigma\alpha}^{\phantom{\dagger}}c_{\sigma \beta}^{\dagger}}
{\vert \eta_\alpha - \eta_\beta \vert^2}
\varphi_{mn}\right]\!\![z_i;w_k;h_1,h_2]
=\Bigg[ \frac{N^2-1}{2}+\frac{3}{2}m(m-N)+\frac{3}{2}n(n-N)\nonumber\\
& &+(n+m)(N-2)-2C_1^2-2C_2-(m-n)\frac{h_1+h_2}{h_1-h_2}+\nonumber\\
& &\Biggl\{-\frac{N-1}{2}\left(\sum_{i=1}^{M_1}\frac{h_1}{h_1-z_i}+
\sum_{k=1}^{M_2}\frac{h_1}{h_1-w_k}\right)\nonumber\\
& &+\sum_{k=1}^{M_2}\frac{h_1^2}{(h_1-w_k)^2}+\frac{h_1}{h_1-h_2}
\left(\sum_{i=1}^{M_1}\frac{h_1}{h_1-z_i}+
\sum_{k=1}^{M_2}\frac{h_1}{h_1-w_k}\right)\nonumber\\
& &+\sum_{i=1}^{M_1}\frac{h_1^2}{(h_1-z_i)^2}
+\sum_{i\neq j}^{M_1}\frac{h_1^2}{(h_1-z_i)(h_1-z_j)}+
\sum_{i=1}^{M_1}\sum_{k=1}^{M_2}\frac{h_1^2}{(h_1-z_i)(h_1-w_k)}\nonumber\\
& &+\sum_{k\neq l}^{M_2}\frac{h_1^2}{(h_1-w_k)(h_1-w_l)}+
2\frac{h_1^2}{(h_1-h_2)^2}\Biggr\}+\{h_1 \leftrightarrow h_2\}\Bigg]
\varphi_{nm}^{\text{ho}},
\end{eqnarray}
where the constants $C_1$ and $C_2$ are defined as above.

As can be readily verified, all non-diagonal terms cancel. Summing up
the diagonal contributions, we obtain the action of $H_{\text{SU(3)}}$
on the auxiliary wave functions $\varphi_{mn}$,
\begin{equation}
\begin{split}
H_{\text{SU(3)}}\varphi_{mn}=&\frac{2\pi^2}{N^2}
\Bigg[\frac{1}{72}(-40+33N-N^3)+\frac{3}{2}m(m-N)+\frac{3}{2}n(n-N)\\
&\hspace{8mm}
+(n+m)(N-2)+2\frac{h_1^2+h_2^2}{(h_1-h_2)^2}-(m-n)\frac{h_1+h_2}{h_1-h_2}
\Bigg]\varphi_{mn}.
\end{split}
\end{equation}
Using \eqref{phidecomposition}, we thus deduce
\begin{eqnarray}
H_{\text{SU(3)}}\Psi_{mn}^{\text{ho}}&=&
-\frac{\pi^2}{36}\left(N+\frac{3}{N}+\frac{4}{N^2} \right)\Psi_{mn}^{\text{ho}}
\nonumber\\
& &+\frac{3\pi^2}{N^2}\left[\left(m-\frac{N+1}{3} \right)m+ 
\left(n-\frac{N+1}{3} \right)n +\frac{m-n}{3}\right]\Psi_{mn}^{\text{ho}}
\nonumber\\
& &+\frac{2\pi^2}{N^2}(m-n)\sum_{\ell=1}^{\lfloor \frac{m-n}{2}\rfloor}
\Psi_{m-\ell,n+\ell}^{\text{ho}}, \label{su3twoholonenergy1}
\end{eqnarray}
where we have used $\frac{x + y}{x - y} (x^{m} y^{n} - x^{n} y^{m}) =
2 \sum_{l = 0}^{m-n} x^{m-l} y^{n+l} - (x^{m} y^{n} + x^{n} y^{m})$
and $\lfloor \; \rfloor$ denotes the floor function, \ie $\lfloor
x\rfloor$ is the largest integer $l\le x$. First, note that the action
of the Hamiltonian on $\Psi_{mn}^{\text{ho}}$ is trigonal, \ie the
``scattering'' in the last line is only to smaller values of $m-n$.
Second, \eqref{su3twoholonenergy1} shows that the states
$\Psi_{mn}^{\text{ho}}$ form a non-orthogonal set, out of which we can
construct an orthogonal basis of eigenfunctions as it is shown in the
following.

\subsubsection{Energy eigenstates}\label{2subsecsu3twoeig}

Using the Ansatz 
\begin{equation}
\ket{\Phi_{mn}^{\text{ho}}}=\sum_{\ell=0}^{\lfloor\frac{m-n}{2}\rfloor}a_{\ell}^{mn}
\ket{\Psi_{m-\ell,n+\ell}^{\text{ho}}},
\label{su3twoenergystates}
\end{equation}
for the diagonalization of \eqref{su3twoholonenergy1}, we obtain the
recursion relation
\begin{equation}
  a_{\ell}^{mn}=-\frac{1}{3\ell\bigl(\ell+m-n-\frac{1}{3}\bigr)}
  \sum_{l=0}^{\ell-1}(n-m-2l)\,a_{l}^{mn},\quad a_{0}^{mn}=1, 
  \label{su3recursion}
\end{equation}
which defines the two-holon energy eigenstates
(\ref{su3twoenergystates}). The corresponding energies are given by
\begin{equation}
E_{mn}^{\text{ho}}=-\frac{\pi^2}{36}\left(N+\frac{3}{N}+\frac{4}{N^2} \right)+
\frac{3\pi^2}{N^2}\left[\left(m-\frac{N+1}{3} \right)m +
\left(n-\frac{N+1}{3} \right)n +\frac{m-n}{3}\right]\!,
\label{su3twofinalenergy} 
\end{equation}
where the momentum quantum numbers are restricted to the interval
\eqref{su3tworestriction} and the total momentum is given by
\eqref{momentumtwoholes}.

The two-holon energies can be rewritten using the one-holon
dispersion \eqref{su3oneholondispersion} as
\begin{equation}
E_{mn}^{\text{ho}}=E_0-\frac{4}{9}\frac{\pi^2}{N^2}
+\epsilon^{\text{ho}}\bigl(p_m^{\text{ho}}\bigr)
+\epsilon^{\text{ho}}\bigl(p_n^{\text{ho}}\bigr), 
\label{su3fractional}
\end{equation}
where we have introduced single-holon momenta according to 
\begin{equation}
p_m^{\text{ho}}=\frac{2\pi}{3}+\frac{2\pi}{N}m, \qquad 
p_n^{\text{ho}}=\frac{2\pi}{3}+\frac{2\pi}{N}\left(n-\frac{1}{3} \right).
\label{su3fractionalmomenta}
\end{equation} 
We will discuss the physical interpretation of this assignment in
Section~\ref{secfrac}.

\section{SU($\bs{n}$) Kuramoto--Yokoyama model}\label{secsun}

In this section we extend our investigations to the SU($n$) KYM. We
will concentrate on stating the results and make only short
remarks on the calculation, since the decisive methods were already
discussed in detail for the SU(3) case.

\subsection{Hamiltonian}\label{subsecsunham}

Consider an underdoped chain with at most one particle per lattice
site carrying an internal SU($n$) quantum number which transforms
according to the fundamental representation $\bs{n}$ of SU($n$).
Starting from the general expression \eqref{ky1} for the SU($n$) KYM,
the Hamiltonian can be rewritten as
\begin{equation}
H_{\text{SU}(n)} =  \frac{2\pi^2}{N^2} 
\sum_{\alpha \neq \beta}^N \frac{1}{ | \eta_\alpha - \eta_\beta |^2 } 
P_{\text{G}} \Biggl[-\frac{1}{2} 
\sum_{\sigma}\Bigl(c_{\alpha \sigma}^\dagger 
c_{\beta \sigma}^{\phantom{\dagger}} 
+ c_{\beta \sigma}^\dagger c_{\alpha \sigma}^{\phantom{\dagger}} \Bigr)
+\frac{1}{2}\sum_{\sigma,\tau}e_{\alpha}^{\sigma \tau}e_{\beta}^{\tau \sigma}
-\frac{n_{\alpha}n_{\beta}}{2}+n_{\alpha}-\frac{1}{2}\Biggr] P_{\text{G}},
\label{kyhamiltoniansun}
\end{equation}
where the summation index $\sigma$ runs over all flavors $1,\ldots,n$,
and the Gutzwiller projector $P_{\text{G}}$ enforces at most single
occupancy on all lattice sites. The model possesses an SU($1|n$)
symmetry generated by the traceless parts of the operators
$J^{ab}=\sum_\alpha a_{\alpha a}^{\dagger}a_{\alpha
  b}^{\phantom{\dagger}}$, where $a_{\alpha a}$ annihilates a particle
of flavor $a$ at site $\eta_\alpha$, as well as a super-Yangian
symmetry~\cite{ha-94prl2887}.

\subsection{Vacuum state}\label{subsecsunvac}

We first consider the state containing no excitations.  We use a
polarized state of particles of flavor $n$ as reference state and
label the coordinates of the particles of flavor $\sigma,
1\le\sigma\le n-1$, by $z_i^\sigma, 1\le i \le M_\sigma$. It can be
shown that the states with wave functions~\cite{Kawakami92prb2}
\begin{equation}
\Psi_0[z_i^\sigma]=\prod_{\sigma=1}^{n-1}
\prod^{M_\sigma}_{i<j}(z_i^\sigma-z_j^\sigma)^2
\prod^{n-1}_{\sigma<\tau}
\prod_{i=1}^{M_\sigma}\prod_{j=1}^{M_\tau}(z_i^\sigma-z_j^\tau)
\prod_{\sigma=1}^{n-1}\prod_{i=1}^{M_\sigma}z_i^\sigma
\label{eq:su3-definitionsunpsi0}
\end{equation}
constitute exact eigenstates~\cite{HaHaldane92} of the Hamiltonian
(\ref{kyhamiltoniansun}).  For $N=nM$, $M_\sigma=M$, \ie at one $n$th
filling, (\ref{eq:su3-definitionsunpsi0}) is the ground state of
(\ref{kyhamiltoniansun}) with energy
\begin{equation}
E_0=-\frac{\pi^2}{12}\left(\frac{n-2}{n}N+\frac{2n-1}{N}\right)\!.
\end{equation}
The momentum is $p=(n-1)\pi M\!\!\mod{2\pi}$, \ie $p=0$ for $n$ odd
and $p=0$ or $p=\pi$ otherwise.

\subsection{Spinon excitations}\label{subsecsunspin}

For $N=nM-1$, localized SU($n$) spinons are represented by the wave
function~\cite{Schuricht-06prb}
\begin{equation}
\Psi_\gamma^{\text{sp}}[z_i^\sigma]=
\prod_{i=1}^{M_1}\bigl(\eta_\gamma-z_i^1\bigr)\,\Psi_0[z_i^\sigma],
\label{eq:su3-localspinon}
\end{equation}
where $M_1=M-1$ and $M_2=\ldots=M_{n-1}=M$. The spinons transform
according to the representation $\bs{\bar{n}}$ under SU($n$)
transformations. Momentum eigenstates are constructed via Fourier
transformation, the spinon momenta are given by
\begin{equation}
p_\nu^{\text{sp}}=\frac{n-1}{n}\pi N-
\frac{2\pi}{N}\left(\nu+\frac{n-1}{2n}\right)\!\!\mod{2\pi},
\label{eq:sunmomenta}
\end{equation}
where the momentum quantum number $\nu$ is restricted to $0\le\nu\le
M_1$. The momenta \eqref{eq:sunmomenta} fill the interval
$[-\frac{\pi}{n},\frac{\pi}{n}]$ for $n$ even and $M$ odd, or the
interval $[\pi-\frac{\pi}{n},\pi+\frac{\pi}{n}]$ otherwise.  The
one-spinon energies are given by
\begin{equation}
E^{\text{sp}}_m=E_0+\frac{n^2-1}{12n}\frac{\pi^2}{N^2}\,+\,
\epsilon^{\text{sp}}\bigl(p_\nu^{\text{sp}}\bigr),
\end{equation}
with \begin{equation}
  \label{eq:epsilonpn}
  \epsilon^{\text{sp}}(p)=\left\{ 
    \begin{array}{l@{\hspace{15pt}}l}
      \displaystyle
      \frac{n}{4}\left(\frac{\pi^2}{n^2}-p^2\right)\!, 
      & \text{if $n$ even and $M$ odd},\\[15pt]
      \displaystyle
      \frac{n}{4}\left(\frac{\pi^2}{n^2}-(p-\pi)^2\right)\!, 
      & \text{otherwise}.
    \end{array}\right.
\end{equation}
SU($n$) spinons obey fractional statistics, the statistical parameter
between spin-polarized spinons is given by $g=(n-1)/n$.

\subsection{One-Holon excitations}\label{subsecsunone}

\begin{figure}[t]
\includegraphics[scale=1]{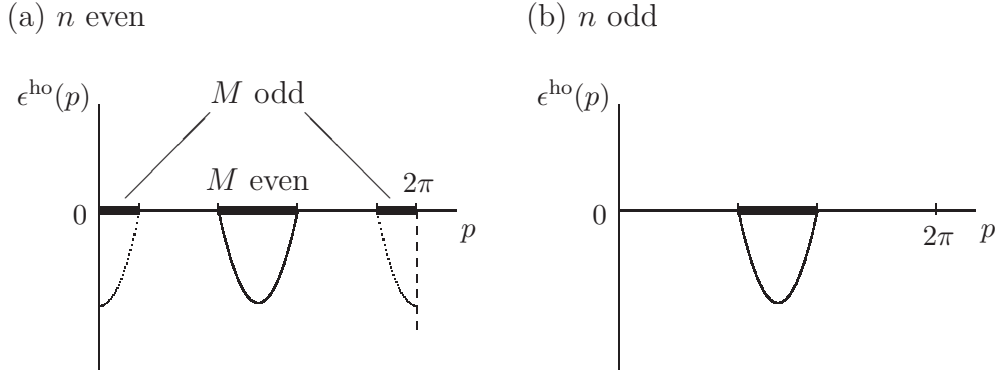}
\caption{SU($n$) holon dispersion. a) $n$ even. The allowed momenta
  fill the interval $[-\frac{\pi}{n},\frac{\pi}{n}]$ for $M$ odd and
  $[\pi-\frac{\pi}{n},\pi+\frac{\pi}{n}]$ for $M$ even. b) $n$ odd.
  The allowed momenta fill the interval
  $[\pi-\frac{\pi}{n},\pi+\frac{\pi}{n}]$.}
 \label{fig:disholonsun}
 \end{figure}

For $N=nM+1$ one can show by a straight-forward generalization of the
SU(3) calculations that the one-holon states represented by the wave
functions
 \begin{equation}
\Psi_{\mu}^{\text{ho}}[z^{\sigma};h]=h^{\mu}\prod_{\sigma=1}^{n-1}
\prod_{i=1}^{M_{\sigma}}\bigl(h-z_{i}^{\sigma}\bigr)\,\Psi_{0}[z^{\sigma}],
\label{sunonehwf}
\end{equation}
are eigenstates of the Hamiltonian \eqref{kyhamiltoniansun}.  In
\eqref{sunonehwf}, $h$ denotes the holon coordinate,
$M_1=\ldots=M_{n-1}=M$, and the momentum quantum number $\mu$ is
restricted to
\begin{equation}
0\le\mu\le\frac{N+n-1}{n}.
\end{equation}
The state corresponding to \eqref{kyhamiltoniansun} is constructed in
analogy to \eqref{su3momonehole}.  The one-holon momenta are given by
\begin{equation}
p_{\mu}^{\text{ho}}=\frac{n-1}{n}\pi N +\frac{2\pi}{N}
\Big(\mu-\frac{n-1}{2n}\Big)\!\!\mod{2\pi},
\end{equation}
which fill the interval $[-\frac{\pi}{n},\frac{\pi}{n}]$ for $n$ even
and $M$ odd, or the interval $[\pi-\frac{\pi}{n},\pi+\frac{\pi}{n}]$
otherwise (either $n$ odd or $M$ even or both).  The one-holon
energies are
\begin{equation}
  E^{\text{ho}}_m=E_0-\frac{n^2-1}{12n}
\frac{\pi^2}{N^2}\,+\,\epsilon^{\text{ho}}\bigl(p_{\mu}^{\text{ho}}\bigr),
\end{equation}
with the single-holon dispersion (see Fig.~\ref{fig:disholonsun})
\begin{equation}
  \label{eq:hoepsilonpn}
  \epsilon^{\text{ho}}(p)=\left\{ 
    \begin{array}{l@{\hspace{15pt}}l}
      \displaystyle
      -\frac{n}{4}\left(\frac{\pi^2}{n^2}-p^2\right)\!, 
      & \text{if $n$ even and $M$ odd},\\[15pt]
      \displaystyle
      -\frac{n}{4}\left(\frac{\pi^2}{n^2}-(p-\pi)^2\right)\!, 
      & \text{otherwise}.
    \end{array}\right.
\end{equation}

\subsection{Two-Holon excitations}\label{subsecsuntwo}

Consider a chain with $N=nM+2$ lattice sites. The two-holon momentum
eigenstates are represented by the wave function
\begin{equation}
\Psi_{\mu\nu}^{\text{ho}}[z_{\sigma};h_1,h_2]=(h_1-h_2)(h_1^{\mu}
h_2^{\nu}+h_1^{\nu}h_2^{\mu})\prod_{\sigma=1}^{n-1}
\prod_{i=1}^{M_{\sigma}}\bigl(h_1-z_{i}^{\sigma}\bigr)
\bigl(h_2-z_{i}^{\sigma}\bigr)\,\Psi_{0}[z_i;w_k], 
\label{suntwowave}
\end{equation}
where the momentum quantum numbers $\mu$ and $\nu$ are restricted to
\begin{equation}
0 \leq \nu \leq \mu \leq \frac{N+n-2}{n}. \label{sunrestriction}
\end{equation}
The total momentum is given by
\begin{equation}
p_{\mu\nu}^{\text{ho}}=\frac{n-1}{n}\pi N+\frac{2\pi}{N}
\left(\mu+\nu-\frac{n-2}{n}\right)\!\!\mod{2\pi}.
\end{equation}
As in the SU(3) case, the momentum eigenstates \eqref{suntwowave} form
a non-orthogonal basis. The two-holon energy eigenstates are obtained
using the Ansatz
\begin{equation}
\ket{\Phi_{\mu\nu}^{\text{ho}}}=\sum_{\lambda=0}^{\lfloor\frac{\mu-\nu}{2}\rfloor}
a_{\lambda}^{\mu\nu}\ket{\Psi_{\mu-\lambda,\nu+\lambda}^{\text{ho}}},
\end{equation}
where the recursion relation for the coefficients
$a_\lambda^{\mu\nu}$'s is found to be
\begin{equation}
a_\lambda^{\mu\nu}=-\frac{1}{n\lambda\bigl(\lambda+\mu-\nu-\frac{1}{n}\bigr)}
\sum_{\kappa=0}^{\lambda-1}(\nu-\mu-2\kappa)a_{\kappa}^{\mu\nu},
\quad a_{0}^{\mu\nu}=1.
\end{equation}
The two-holon energies are given by
\begin{eqnarray}
E_{\mu\nu}^{\text{ho}}&=&-\frac{\pi^2}{12n}\left(\bigl(n-2\bigr)N+
\bigl(2n^2-13n+24\bigr)\frac{1}{N}
-4\bigl(n^2-6n+8\bigr)\frac{1}{N^2}\right)\nonumber \\[2mm]
&&+\frac{n\pi^2}{N^2}\bigg[\left(\mu-\frac{N+n-2}{n}\right)\mu+
\left(\nu-\frac{N+n-2}{n}\right)\nu+\frac{\mu-\nu}{n} \bigg]. 
\label{finalsunenergyho}
\end{eqnarray}
Using the single-holon dispersions \eqref{eq:hoepsilonpn}, the energy
eigenvalues of~\eqref{finalsunenergyho} can be rewritten as
\begin{equation}
E_{\mu\nu}^{\text{ho}}=E_0-\frac{n^2-1}{6n}\frac{\pi^2}{N^2}
+\epsilon^{\text{ho}}\bigl(p_\mu^{\text{ho}}\bigr)
+\epsilon^{\text{ho}}\bigl(p_\nu^{\text{ho}}\bigr), 
\label{sunfractional}
\end{equation}
where we have introduced single-holon momenta according to 
\begin{equation}
p_\mu^{\text{ho}}=-\frac{\pi}{n}+\frac{2\pi}{N}\left(\mu-\frac{n-3}{2n}\right), 
\qquad 
p_\nu^{\text{ho}}=-\frac{\pi}{n}+\frac{2\pi}{N}\left(\nu-\frac{n-1}{2n}\right),
\label{sunfractionalmomenta}
\end{equation} 
and restricted ourselves to momenta
$-\frac{\pi}{n}\le p_\nu^{\text{ho}}\le p_\mu^{\text{ho}}\le\frac{\pi}{n}$
for simplicity.

\section{Fractional Statistics}\label{secfrac}

Fractional statistics in one dimension was originally introduced by
Haldane~\cite{Haldane91prl2} in terms of non-trivial state counting
rules.  Recently, it was realized that the fractional statistics of
spinons and holons in the KYM manifests itself also in specific
quantization rules for the individual spinon and holon
momenta~\cite{Greiter-05prb,Greiter06,TSG}.  Here we apply this
interpretation to the holon excitations of the SU($n$) KYM.

First, consider holons in the SU(3) KYM. As we have seen in
\eqref{su3fractional}, the two-holon energies are simply given by the
sum of the kinetic energies of the individual holons (and the ground
state energy). This shows that the holons in the SU(3) KYM are free, which
is supported by conclusions drawn from the asymptotic Bethe
Ansatz~\cite{Essler95}.  Furthermore, the momentum spacing between the
individual holon momenta 
in \eqref{su3fractionalmomenta} is
\begin{equation}
  \label{eq:pmminuspn3}
  p_m-p_n=\frac{2\pi}{N}\left(\frac{1}{3}+\ell\right),
  \quad \ell\in\mathbb{N}_0,
\end{equation}
which reflects the fractional statistics of the holons with
statistical parameter $g=1/3$. This result is consistent with
conclusions reached by Kuramoto and Kato~\cite{kuramoto,kato} from
thermodynamics, and by Arikawa, Yamamoto, Saiga, and
Kuramoto~\cite{Arikawa3} from the charge dynamics of the model.

For holons in the SU($n$) KYM the situation is similar. From
\eqref{sunfractional} we deduce that the holons are free, whereas the
momentum spacings
\begin{equation}
  \label{eq:pmminuspnn}
  p_m-p_n=\frac{2\pi}{N}\left(\frac{1}{n}+\ell\right),
  \quad \ell\in\mathbb{N}_0,
\end{equation}
obtained
from \eqref{sunfractionalmomenta} show that holons in the SU($n$) KYM
obey fractional statistics with statistical parameter $g=1/n$. 

Derived in the context of the KYM, this result has implications for
SU($n$) spin chains in general.  In the KYM, where the holons are free
in the sense that they only interact through their fractional
statistics, the individual holon momenta are good quantum
numbers. They assume fractionally spaced values, which for two
holons are given by \eqref{eq:pmminuspnn}.  As the statistics of the
holons is a quantum invariant and as such independent of the details
of the model, the fractional spacings are of universal validity as
well. If we were to supplement the KYM by a potential interaction
between the holons, this interaction would introduce scattering matrix
elements between the exact eigenstates we obtained and labeled
according to their fractionally spaced single particle momenta.  These
momenta would hence no longer constitute good quantum numbers.  The
new eigenstates would be superpositions of states with different
single particle momenta, which individually, however, would still
possess the fractionally shifted values. The effect of the interaction
would hence be to turn the integer $\ell$ on the right-hand side
of~\eqref{eq:pmminuspnn} into a superposition of integers, while
leaving the fractional momentum spacing $2\pi / N n$ unchanged.

Note that regardless of $n$, the sum of the statistical parameters of
spinons and holons always equals the fermionic value one,
\begin{equation}
g_{\text{sp}}+g_{\text{ho}}=\frac{n-1}{n}+\frac{1}{n}=1,
\label{eq:gsum}
\end{equation} 
a result consistent with the concept of spin-charge separation 
characteristic of these models.

Finally, as models with SU($n$) symmetry in general are frequently
studied because of simplifying features, it is suggestive to ask
whether the large-$n$ limit deserves special attention in the model we
have studied here as well.  Briefly, the answer is no.  No part of our
calculation simplifies in this limit, as we obtain terms similar to
the ones encountered above regardless of the value of $n$.  In the
limit $n\rightarrow\infty$, $g\rightarrow 0$ implies that the
exclusion statistics between holons tends towards bosons.  This does
not mean, however, that the holons in this limit behave like free
bosons, but rather that their momentum spacings shrink with the $n$th
part of the Brillouin zone they are confined to.

\section{Conclusions}\label{secconc}

In conclusion, we have constructed the explicit wave functions of the
one- and two-holon excitations of the SU($n$) KYM and derived their
exact energies.  The holons are non-interacting or free, but obey
fractional statistics with parameter $g=1/n$, which manifests itself
in the quantization of the single holon momenta, which is a general
feature of fractional charge excitations in SU($n$) spin chains.

\section*{ACKNOWLEDGMENTS}

RT was supported by a PhD scholarship of the Studienstiftung des
deutschen Volkes, and DS by the Center for
Functional Nanostructures (CFN) Karlsruhe.

\appendix
\section{Gell-Mann matrices}\label{app:conventions}

The Gell-Mann matrices are explicitly given by~\cite{Cornwell84vol2}
\begin{displaymath}
\begin{split}
&\lambda^1=\left(\begin{array}{ccc}0&1&0\\1&0&0\\0&0&0\end{array}
\right)\!\!,\quad
\lambda^2 =\left(\begin{array}{ccc}0&-i&0\\i&0&0\\0&0&0\end{array}
\right)\!\!,\quad
\lambda^3 =\left(\begin{array}{ccc}1&0&0\\0&-1&0\\0&0&0\end{array}
\right)\!\!,\\[3mm]
&\lambda^4=\left(\begin{array}{ccc}0&0&1\\0&0&0\\1&0&0\end{array}
\right)\!\!,\quad
\lambda^5 =\left(\begin{array}{ccc}0&0&-i\\0&0&0\\i&0&0\end{array}
\right)\!\!,\quad
\lambda^6 =\left(\begin{array}{ccc}0&0&0\\0&0&1\\0&1&0\end{array}
\right)\!\!,\\[3mm]
&\lambda^7=\left(\begin{array}{ccc}0&0&0\\0&0&-i\\0&i&0\end{array}
\right)\!\!,\quad
\lambda^8 =\frac{1}{\sqrt{3}}
\left(\begin{array}{ccc}1&0&0\\0&1&0\\0&0&-2\end{array}\right)\!\!.
\end{split}
\end{displaymath}
They are normalized as
$\mathrm{tr}\left(\lambda^a\lambda^b\right)=2\delta_{ab}$ and satisfy
the commutation relations
$\comm{\lambda^a}{\lambda^b}=2f^{abc}\lambda^c.$ The structure
constants $f^{abc}$ are totally antisymmetric and obey Jacobi's
identity
\begin{displaymath}
f^{abc}f^{cde}+f^{bdc}f^{cae}+f^{dac}f^{cbe}=0.
\end{displaymath} 
Explicitly, the non-vanishing structure constants are given by
$f^{123}=i$, $f^{147}=f^{246}=f^{257}=f^{345}=-f^{156}=-f^{367}=i/2$,
$f^{458}=f^{678}=i\sqrt{3}/2$, and 45 others obtained by permutations
of the indices.  

The SU(3) spin operators can be expressed in terms of the colorflip
operators and the charge occupation operator as
\begin{displaymath}
\bs{J}_\alpha\!\cdot\!\bs{J}_\beta\equiv
\sum_{a=1}^8 J^a_\alpha J^a_\beta=
\frac{1}{2}\sum_{\sigma\tau}^3\,e_\alpha^{\sigma\tau}\,e_\beta^{\tau\sigma} 
-\frac{1}{6}n_\alpha n_\beta.
\end{displaymath}

\section{Useful formulas}\label{app:formulas}

For derivations see for example~\cite{BGLtwospinon,Schuricht-06prb}.

\begin{enumerate}
\item 
\begin{equation}
\eta_\alpha^N=1,\quad
\sum_{\alpha=1}^N\eta_\alpha^m=N\,\delta_{0m}, \quad
\prod_{\alpha=1}^N\eta_\alpha=(-1)^{N-1}.
\label{eq:appB1}
\end{equation}

\item 
\begin{equation}
\frac{1}{\vert \eta_{\alpha}-\eta_{\beta}\vert^2}=
-\frac{\eta_{\alpha}\eta_{\beta}}{(\eta_{\alpha}-\eta_{\beta})^2}.
\label{eq:hs1/vertetavert}
\end{equation}

\item
\begin{equation}
\prod_{\alpha=1}^N(\eta-\eta_\alpha)=\eta^N-1.
\end{equation}

\item
\begin{equation}
\prod_{\beta \neq \alpha}^N(\eta_{\beta}-\eta_{\alpha})=
\lim_{\eta \rightarrow \eta_{\alpha}} \frac{\eta^{N}-1}{\eta-\eta_{\alpha}}=
\frac{N}{\eta_{\alpha}}.
\label{eq:app-holon1}
\end{equation}

\item
\begin{equation}
  \sum_{\alpha=1}^{N-1}\frac{\eta_\alpha^m}{\eta_\alpha -1}
  =\frac{N+1}{2}-m,
  \quad 1\le m \le N.
  \label{eq:app-hsfouriersum1}
\end{equation}

\item
\begin{equation}
  \sum_{\alpha=1}^{N-1}
  \frac{\eta_\alpha^m}{\vert\eta_\alpha -1\vert^2}=
  -\sum_{\alpha=1}^{N-1}
  \frac{\eta_\alpha^{m+1}}{(\eta_\alpha -1)^2}=
  \frac{N^2-1}{12}+\frac{m(m-N)}{2},
  \quad 0\le m \le N.
\label{eq:app-hsfouriersum2}
\end{equation}

\end{enumerate}

\section{Representation of wave functions}\label{appsecrep}

It is shown that the wave functions can, up to a minus sign, be
expressed by any two sets of color variables. First, the wave function
of the vacuum state \eqref{eq:su3-definitionpsi0} can be rewritten
using green ($u$) variables as
\begin{eqnarray}
\Psi_{0}[z_i;w_k]&=&(-1)^{M_1\frac{M_1-1}{2}}\prod_{i\neq j}^{M_{1}}(z_i-z_j)
\prod_{k<l}^{M_{2}}(w_{k}-w_{l})^2\prod_{i=1}^{M_{1}}\prod_{k=1}^{M_2}
(z_i-w_k)\prod_{i=1}^{M_1}z_i\prod_{k=1}^{M_2}w_k\nonumber\\
&=&(-1)^{M_1\frac{M_1-1}{2}}(-1)^{M_1M_2}
\frac{\prod_{i=1}^{M_{1}}z_{i}\prod_{k=1}^{M_{2}}w_{k}
\prod_{k<l}^{M_{2}}(w_{k}-w_{l})^2\prod_{i=1}^{M_{1}}
\frac{N}{z_{i}}}{\prod_{i=1}^{M_{1}}\prod_{s=1}^{M_{3}}(u_{s}-z_{i})}\nonumber\\
  &=&(-1)^{M_1\frac{M_1-1}{2}}(-1)^{M_1M_2}\frac{\prod_{k=1}^{M_{2}}w_{k}
\prod_{k<l}^{M_{2}}(w_{k}-w_{l})^2N^{M_{1}}}
{\prod_{i=1}^{M_{1}}\prod_{s=1}^{M_{3}}(u_{s}-z_{i})},
\label{groundid1}
\end{eqnarray}
where we have used \eqref{eq:app-holon1}. Accordingly, if we express
$\Psi_0$ in terms of green and red variables, we find
\begin{eqnarray}
\Psi_{0}[u_{s};w_{k}]
&=&(-1)^{M_3\frac{M_3-1}{2}}\prod_{s\neq t}^{M_{3}}(u_{s}-u_{t})
\prod_{k<l}^{M_{2}}(w_{k}-w_{l})^2\prod_{s=1}^{M_{3}}
\prod_{k=1}^{M_2}(u_s-w_k)\prod_{s=1}^{M_3}u_s\prod_{k=1}^{M_2}w_k\nonumber\\
&=&(-1)^{M_3\frac{M_3-1}{2}}(-1)^{M_2M_3}
\frac{\prod_{s=1}^{M_{3}}u_{s}\prod_{k=1}^{M_{2}}w_{k}
\prod_{k<l}^{M_{2}}(w_{k}-w_{l})^2
\prod_{s=1}^{M_{3}}\frac{N}{u_{s}}}{\prod_{i=1}^{M_{1}}
\prod_{s=1}^{M_{3}}(u_{s}-z_{i})} \nonumber\\
&=&(-1)^{M_3\frac{M_3-1}{2}}(-1)^{M_2M_3}(-1)^{M_{1}M_{3}}
\frac{\prod_{k=1}^{M_{2}}w_{k}\prod_{k<l}^{M_{2}}
(w_{k}-w_{l})^2N^{M_{3}}}{\prod_{i=1}^{M_{1}}
\prod_{s=1}^{M_{3}}(u_{s}-z_{i})} \nonumber\\
&=&(-1)^{M^2}\Psi_{0}[z_i;w_k],
\label{groundid2}
\end{eqnarray}
where we again used \eqref{eq:app-holon1}, and finally set
$M_1=M_2=M_3=M$.

The same line of argument can be applied to the one-holon wave
functions \eqref{su3waveonehole},
\begin{equation}
\Psi_{m}^{\text{ho}}[z_i;w_k;h]
=(-1)^{M_1\frac{M_1-1}{2}}(-1)^{M_1M_2}h^n\frac{\prod_{k=1}^{M_{2}}
(h-w_{k})\prod_{k=1}^{M_{2}}w_{k}\prod_{\substack{k,l\\k<l}}^{M_{2}}
(w_{k}-w_{l})^2N^{M_{1}}}{\prod_{i=1}^{M_{1}}
\prod_{s=1}^{M_{3}}(u_{s}-z_{i})}, \label{hwfid1}
\end{equation}
whereas starting with green and red variables yields
\begin{eqnarray}
& &\hspace{-10mm}\Psi_{m}^{\text{ho}}[u_s;w_k;h]\nonumber\\
&=&(-1)^{M_3\frac{M_3-1}{2}}(-1)^{M_1M_3}(-1)^{M_{2}M_{3}}h^n 
\frac{\prod_{k=1}^{M_{2}}(h-w_{k})\prod_{k=1}^{M_{2}}w_{k}
\prod_{\substack{k,l\\k<l}}^{M_{2}}(w_{k}-w_{l})^2N^{M_{3}}}
{\prod_{i=1}^{M_{1}}\prod_{s=1}^{M_{3}}(u_{s}-z_{i})} \nonumber\\
&=&(-1)^{M^2}\Psi_{m}^{\text{ho}}[z_{i};w_{k};h].
\label{hwfid2}
\end{eqnarray}
In the same way we find for the two-holon wave functions
\eqref{su3wavetwoholes}
\begin{equation}
\Psi_{mn}^{\text{ho}}[z_i;w_k;h_1,h_2]=(-1)^{M^2}\Psi_{mn}^{\text{ho}}
[u_s;w_k;h_1,h_2].
\end{equation}
Thus, the holon wave functions can be expressed by any two sets of
color indices. All statements generalize to SU($n$).

\section{$\bs{B}$-Series}\label{app:bseries}

The $B$-series is defined as 
\begin{equation}
B_\ell^m=-\sum_{\alpha=1}^{N-1}\eta_{\alpha}^{m+1}(\eta_{\alpha}-1)^{\ell-2}, 
\label{eq:app-su3twoseries}
\end{equation}
where $0 \leq \ell \leq 2(N-1)/3$. Now, $B_0^m$ equals
\eqref{eq:app-hsfouriersum2}, $B_1^m=m-(N-1)/2$ for $0 \le m <N$ by
(\ref{eq:app-hsfouriersum2}), and $B_2^m=1$ for $0\le m<N+1$ by
\eqref{eq:appB1}. Furthermore, for $3\le\ell\leq 2(N-1)/3$ we find
\begin{equation}
B_\ell^m=\left\{
\begin{aligned} 
0,\quad&\mathrm{for}&\quad 0 \le m \le \frac{N+2}{3}, \\
N{\ell-2 \choose N-m-1},\quad&\mathrm{for}&\quad\frac{N+2}{3} < m \le N.
 \end{aligned} \right. 
 \label{eq:app-su3twoseries4}
\end{equation}
{\em Proof:} 
\begin{eqnarray*}
B_\ell^m&=&-\sum_{\alpha=1}^{N-1} \eta_\alpha^{m+1}\sum_{k=0}^{\ell-2}
{\ell-2 \choose k}(-1)^{\ell-k-2}\eta_\alpha^k\\
&=&-\sum_{k=0}^{\ell-2}{\ell-2 \choose k}(-1)^{\ell-k}
\left(1-\sum_{\alpha=1}^N \eta_\alpha^{m+k+1}\right)\\
&=&-\sum_{k=0}^{\ell-2} {\ell-2 \choose k}(-1)^{\ell-k}
\bigl(1-N\delta_{m,N-k-1}\bigr).
\end{eqnarray*}
Thus, for $0\le m\le (N+2)/3$, $B_\ell^m$ vanishes, as the sums of the
binomial coefficients of even sites and odd sites equal each other.
For $(N+2)/3<m$, however, $B_\ell^m\neq 0$, and thus the Taylor
expansion appearing in the calculations of the charge kinetic terms
contains higher order derivatives.

The remaining constants are deduced from $A_\ell=B_\ell^1$,
$C_1=B_1^{N-1}$, and $C_2=-B_0^{N-1}$.

\section{Derivative identities}\label{appsecderive}

If one holon is present, we use for the simplification of the charge
kinetic terms
\begin{eqnarray}
\sum_{s=1}^{M_{3}}\frac{h}{h-u_{s}}&=&
\sum_{\alpha\neq h}^{N}\frac{h}{h-\eta_{\alpha}}-\sum_{i=1}^{M_{1}}
\frac{h}{h-z_{i}}-\sum_{k=1}^{M_{2}}\frac{h}{h-w_{k}}\nonumber\\
&=&\frac{N-1}{2}-\sum_{i=1}^{M_{1}}
\frac{h}{h-z_{i}}-\sum_{k=1}^{M_{2}}\frac{h}{h-w_{k}},
\label{su3A}
\end{eqnarray}
as well as
\begin{eqnarray}
\sum_{s \neq t}^{M_3}\frac{h^2}{(h-u_{s})(h-u_{t})}&=&
\sum_{s,t}^{M_3}\frac{h^2}{(h-u_{s})(h-u_{t})}-
\sum_{s=1}^{M_{3}}\frac{h^2}{(h-u_s)^2} \nonumber \\
&=&C_1^2-C_2+\sum_{i=1}^{M_{1}}\frac{2h^2}{(h-z_{i})^2}+\sum_{k=1}^{M_{2}}
\frac{2h^2}{(h-w_{k})^2} \nonumber \\*
& &-C_1\sum_{i=1}^{M_{1}}\frac{2h}{h-z_{i}}-C_1\sum_{k=1}^{M_{2}}
\frac{2h}{h-w_{k}}+\sum_{i \neq j}^{M_1}\frac{h^2}{(h-z_i)(h-z_j)}\nonumber \\*
& &+\sum_{k \neq l}^{M_2}\frac{h^2}{(h-w_{k})
(h-w_{l})}+\sum_{i=1}^{M_{1}}\sum_{k=1}^{M_{2}}
\frac{2h^2}{(h-z_{i})(h-w_{k})},
\label{su3derivativeid1}
\end{eqnarray}
with the constants $C_{1}$ and $C_2$ as above.

For the two-holon case, we apply the identity
\begin{eqnarray}
\sum_{s \neq t} \frac{h_1^2}{(h_1-u_s)(h_1-u_t)}
&=&-C_2+\sum_{i=1}^{M_1}\frac{h_1^2}{(h_1-z_i)^2}
+\sum_{k=1}^{M_2}\frac{h_1^2}{(h_1-w_k)^2}+\frac{h_1^2}{(h_1-h_2)^2}\nonumber\\
& &+\left(C_1-\sum_{i=1}^{M_1}\frac{h_1}{h_1-z_i}
-\sum_{k=1}^{M_2}\frac{h_1}{h_1-w_k}-\frac{h_1}{h_1-h_2}\right)\nonumber \\
& &\quad\times\left(C_1-\sum_{j=1}^{M_1}\frac{h_1}{h_1-z_j}
-\sum_{l=1}^{M_2}\frac{h_1}{h_1-w_l}-\frac{h_1}{h_1-h_2}\right)\nonumber\\
&=&-C_2+C_1^2+\sum_{i\neq j}\frac{h_1^2}{(h_1-z_i)(h_1-z_j)}+\sum_{k \neq l}
\frac{h_1^2}{(h_1-w_k)(h_1-w_l)}\nonumber\\
& &+\sum_{i=1}^{M_1}\frac{2h_1^2}{(h_1-z_i)^2}+
\sum_{k=1}^{M_2}\frac{2h_1^2}{(h_1-w_k)^2}+
\frac{2h_1^2}{(h_1-h_2)^2}-C_1\frac{2h_1}{h_1-h_2}\nonumber\\
& &+\sum_{i=1}^{M_1}\sum_{k=1}^{M_2}\frac{2h_1^2}{(h_1-z_i)(h_1-w_k)}-
C_1\sum_{i=1}^{M_1}\frac{2h_1}{h_1-z_i}-
C_1\sum_{k=1}^{M_2}\frac{2h_1}{h_1-w_k}\nonumber\\
& &+\frac{2h_1}{h_1-h_2}\left(\sum_{i=1}^{M_1}\frac{h_1}{h_1-z_i}+
\sum_{k=1}^{M_2}\frac{h_1}{h_1-w_k}\right),
\end{eqnarray}
and the similar result for $h_1\leftrightarrow h_2$.  All identities
presented above directly generalize to SU($n$).

\end{document}